\documentclass[twocolumn]{aastex631}

\usepackage[version=4]{mhchem}
\let\tablenum\relax
\usepackage{siunitx}
\usepackage{amsmath}
\usepackage{chemformula}
\usepackage{stackengine}
\usepackage{makecell}
                              
\graphicspath{{./}{figures/}}

\begin{document}

\title{Probing the kinematics of FU Orionis objects through high-resolution near-infrared spectroscopy}

\author[0000-0002-9535-7453]{Ellen Lee}
\affiliation{Institute for Astronomy, University of Hawai'i at Manoa, 640 N. Aohoku Place, Hilo, HI 96720, USA}

\author[0000-0002-8293-1428]{Michael S. Connelley}
\affiliation{Institute for Astronomy, University of Hawai'i at Manoa, 640 N. Aohoku Place, Hilo, HI 96720, USA}

\begin{abstract}
FU Orionis (FUor) objects are thought to be described by a steady-state Keplerian disk. However, the characteristic double-peaked Keplerian line profile is not readily seen in most near-infrared spectra of FUors. In this paper, we measure the near-infrared line profiles of 15 FUors and FUor-like objects by convolving model cool atmosphere spectra with a linear combination of Gaussians. The models are fit to high-resolution spectra obtained with iSHELL on the NASA Infrared Telescope Facility (IRTF). Five of the targets are found to have double-peaked line profiles in K-band, which can also be fitted by a Keplerian line profile. For eight targets that were also observed in J-band, we find that the line profiles are well-correlated to what is observed in K-band, but the linewidth does not clearly appear to decrease with wavelength. We find that a double-peaked line profile can be difficult to see for several reasons, which include blending with extraneous molecular features and potential absorption from a disk wind or infalling material. The CO lines in M-band are morphologically different from their counterparts in K-band, so they are probably of a different origin.
\end{abstract}

\section{Introduction} \label{sec:intro}
FUors are a rare type of outbursting young star characterized by about five magnitudes of increase in brightness that lasts several decades \citep{1966VA......8..109H,1977ApJ...217..693H}. In cases when the outburst has not been observed, FUor-like objects may be classified spectroscopically in the near-infrared due to their unique spectral features; notably, a lack of emission lines, strong CO absorption at K-band, deep HeI absorption around \qty{1.083}{\micro\meter}, and triangular bands of water vapor at H-band, along with an abundance of other molecular absorption bands. The spectra of FUors are relatively stable and do not exhibit drastic changes over decade timescales \citep{connelley2018fuor}. FUors also exhibit a gradual transition toward later spectral types at longer wavelengths. Rare instances of pre-outburst spectra indicate that FUors may have been T Tauri stars \citep{1977ApJ...217..693H,2018AJ....156...25R,2025ApJ...994L..43H}. It is unknown whether the post-outburst spectrum of a FUor will return to this state.

The mechanism that produces FUors is not well understood. Although others have suggested accretion as a source for large outbursts \citep{1976IAUS...73...75P}, \cite{hartmann1985nature} and \cite{1988ApJ...325..231K} first developed the Keplerian accretion disk model that is commonly used to model FUors post-outburst. There are several lines of evidence to support the Keplerian disk model:
\begin{enumerate}
    \item{At optical wavelengths, absorption lines appear to be double-peaked.}
    \item{The optical linewidth is broader than what is measured at the near-infrared, where the cooler, farther regions of the disk dominate the spectrum.}
    \item The observed trend of later spectral types at longer wavelengths can be explained by a self-luminous disk with a radial temperature profile of $T_{eff}\sim r^{-0.75}$.
\end{enumerate}

The characteristic double-peaked line profile is one of the most striking features of Keplerian disks, but it is not immediately apparent in high-resolution measurements of the CO lines beginning around \qty{2.294}{\micro\meter}. They instead appear to be Gaussian or box-shaped; in cases where double-peaked line profiles are argued to be observed, the spectra do not appear to be well matched to a Keplerian disk profile \citep{1988ApJ...325..231K, hartmann1996fu, 2004ApJ...609..906H, 2008AJ....135.1421G, 2020ApJ...900...36P}. \cite{2003ApJ...595..384H} and \cite{2008AJ....136..676P} also comment on the subtlety (or absence) of a double-peaked line structure in several absorption lines at optical wavelengths as well as the lack of a clearly discernible wavelength dependence in the linewidth. More recent, sophisticated accretion disk models are significantly better at modeling the observed spectra of FUors, but do not always fully reproduce the line structure in the near-infrared \citep{2025ApJ...993...38C}. In this paper, we present a high-resolution spectroscopic survey of FUors with the goal of verifying whether their near-infrared line profiles match a Keplerian disk model.

\section{Observations} \label{sec:observations}
We performed our observations with iSHELL at NASA IRTF \citep{rayner2022ishell} between August 25, 2017 and August 24, 2024. We used the 0\farcs75 slit to obtain resolving powers of $R\sim50,000$. Our sample consists of the majority of FUors and FUor-like objects that are bright enough to be observed by iSHELL. We observed a total of 15 objects using the K2 mode of iSHELL (2.09 - \qty{2.38}{\micro\meter}), of which nine were bright enough to also be observed with the J3 mode (1.26 - \qty{1.36}{\micro\meter}). Five objects were also observed with the M2 mode (4.52 - \qty{5.25}{\micro\meter}) along with HBC 722 which was only observed in M-band. Details of the observations are provided in Tables~\ref{tab:obslog} and \ref{tab:obslog-m2}.

%% The values (usually only l,r and c) in the last part of
%% \begin{deluxetable}{} command tell LaTeX how many columns
%% there are and how to align them.
\begin{deluxetable*}{lccccccccc}[h]
\tabletypesize{\scriptsize}
%% Keep a portrait orientation

%% Over-ride the default font size
%% Use Default (12pt)

%% Use \tablewidth{?pt} to over-ride the default table width.
%% If you are unhappy with the default look at the end of the
%% *.log file to see what the default was set at before adjusting
%% this value.

%% This is the title of the table.
\label{tab:obslog}
\tablecaption{Observing log for J- and K-band. Integration times are in minutes. For objects that were observed more than once, we used data from the dates that were closest to when the object was observed in the other wavelength band (in bold).}

%% This command over-rides LaTeX's natural table count
%% and replaces it with this number.  LaTeX will increment 
%% all other tables after this table based on this number
%%\tablenum{1}

%% The \tablehead gives provides the column headers.  It
%% is currently set up so that the column labels are on the
%% top line and the units surrounded by ()s are in the 
%% bottom line.  You may add more header information by writing
%% another line between these lines. For each column that requries
%% extra information be sure to include a \colhead{text} command
%% and remember to end any extra lines with \\ and include the 
%% correct number of &s.
\tablehead{
  \colhead{} & \colhead{} & \colhead{} & \colhead{} & \multicolumn{3}{c}{J}  & \multicolumn{3}{c}{K}\\ 
  \colhead{Object} & \colhead{Alt. Name} & \colhead{RA (J2000)} & \colhead{Dec (J2000)} & \colhead{mag\tablenotemark{a}} & \colhead{i. time} & \colhead{UT date} & \colhead{mag\tablenotemark{a}} & \colhead{i. time} & \colhead{UT date}
}

%% All data must appear between the \startdata and \enddata commands
\startdata
RNO 1b & & 00:36:46.00 & +63:28:52.9 & 10.7 & 60 & 201120 & 7.8 & 15 & 201021 \\
V582 Aur & & 05:25:51.98 & +34:52:30.0 & &  &  & 7.7 & 50 & 201119 \\
Haro 5a/6a & IRAS 05329-0505 & 05:35:26.75 & -05:03:55.1 &  &  &  & 9.9\tablenotemark{b} & 100 & 201119  \\
PR Ori B & & 05:36:25.99 & -06:17:32.6 & 8.6 & 40 & 241015 & 8.2 & 15 & 241015  \\
V883 Ori & & 05:38:18.10 & -07:02:25.9 & 9.3 & 60 & 201119 & 5.2 & 40 & 201119  \\
V2775 Ori & & 05:42:48.49 & -08:16:34.7 & &  &  & 10.0 & 35 & 201120  \\
V960 Mon & & 06:59:31.59 & -04:05:27.8 & &  &  & 8.3 & 26 & 180114 \\
V900 Mon & & 06:57:22.22 & -08:23:17.7 & &  &  & 8.3 & 10 & 180115 \\
FU Ori\tablenotemark{c} & & 05:45:22.36 & +09:04:12.4 & 6.5 & 10 & 160925, \textbf{201120} & 5.2 & 5 & 160918, \textbf{201120} \\
V646 Pup & BBW 76 & 07:50:35.60 & -33:06:23.9 & 9.2 & 25 & 201120 & 7.8 & 25 & 201120 \\
Parsamian 21 & & 19:29:00.86 & +09:38:42.9 & 11.2 & 60 & 201120 & 9.8 & 50 & 170825 \\
V1515 Cyg & & 20:23:48.02 & +42:12:25.8 & 8.9 & 40 & 201023 & 7.4 & 25, 12 & 170825, \textbf{201023} \\
V2494 Cyg &  & 20:58:21.09 & +52:29:27.7 & &  &  & 8.3 & 30 & 201119\\
%V1057 Cyg\tablenotemark{d} & & 20:58:53.73 & +44:15:28.4 & 7.0 & 15 & 240824 & 6.2 & 8, 9 & 170825, 240824 \\
V1735 Cyg & & 21:47:20.66 & +47:32:03.6 & 9.9 & 25, 40 & 170825, \textbf{201021} & 7.0 & 20 & 170826, \textbf{201021} \\
V733 Cep & & 22:53:33.26 & +62:32:23.6 & 10.8 & 60 & 201119 & 8.2 & 20 & 201023 \\
\enddata

%% Include any \tablenotetext{key}{text}, \tablerefs{ref list},
%% or \tablecomments{text} between the \enddata and 
%% \end{deluxetable} commands

%% No \tablecomments indicated
\tablenotetext{a}{J- and K-band magnitudes are from 2MASS \citep{2003yCat.2246....0C}.}
\tablenotetext{b}{The K-band magnitude of Haro 5a/6a is measured by \cite{connelley2018fuor}.}
\tablenotetext{c}{2016 data of FU Ori were measured during the commissioning of iSHELL and are not stored in the IRTF data archive. Additional J-, H-, and K-band from 2016 are available upon request.}
%\tablenotetext{d}{Due to V1057 Cyg's unique spectrum and variability, it is excluded from our line profile analysis.}
%% No \tablerefs indicated
\end{deluxetable*}

%% The values (usually only l,r and c) in the last part of
%% \begin{deluxetable}{} command tell LaTeX how many columns
%% there are and how to align them.
\begin{deluxetable*}{lccccccccc}[h]
\tabletypesize{\scriptsize}
%% Keep a portrait orientation

%% Over-ride the default font size
%% Use Default (12pt)

%% Use \tablewidth{?pt} to over-ride the default table width.
%% If you are unhappy with the default look at the end of the
%% *.log file to see what the default was set at before adjusting
%% this value.

%% This is the title of the table.
\label{tab:obslog-m2}
\tablecaption{Observing log for M-band. Integration times are in minutes.}

%% This command over-rides LaTeX's natural table count
%% and replaces it with this number.  LaTeX will increment 
%% all other tables after this table based on this number
%%\tablenum{1}

%% The \tablehead gives provides the column headers.  It
%% is currently set up so that the column labels are on the
%% top line and the units surrounded by ()s are in the 
%% bottom line.  You may add more header information by writing
%% another line between these lines. For each column that requries
%% extra information be sure to include a \colhead{text} command
%% and remember to end any extra lines with \\ and include the 
%% correct number of &s.
\tablehead{
  \colhead{Object} & \colhead{Alt. Name} & \colhead{RA (J2000)} & \colhead{Dec (J2000)} & \colhead{mag\tablenotemark{a}} & \colhead{i. time} & \colhead{UT date}
}

%% All data must appear between the \startdata and \enddata commands
\startdata
V883 Ori & & 05:38:18.10 & -07:02:25.9 & 1.5 & 15 & 240326 \\
FU Ori & & 05:45:22.36 & +09:04:12.4 & 3.7 & 58 & 240326 \\
V1515 Cyg & & 20:23:48.02 & +42:12:25.8 & 6.0 & 59 & 240714 \\
HBC 722 & V2493 Cyg & 20:58:17.03 & +45:53:43.3 & 3.7 & 118 & 240717 \\
V2494 Cyg & & 20:58:21.09 & +52:29:27.7 & 3.7 & 59 & 240714 \\
%V1057 Cyg & & 20:58:53.73 & +44:15:28.4 & ? & 59 & 240714 \\
V1735 Cyg & & 21:47:20.66 & +47:32:03.6 & 5.1 & 87 & 240717 \\
\enddata

%% Include any \tablenotetext{key}{text}, \tablerefs{ref list},
%% or \tablecomments{text} between the \enddata and 
%% \end{deluxetable} commands

%% No \tablecomments indicated
\tablenotetext{a}{The ALLWISE W2 magnitude ($\lambda$=\qty{4.6}{\micro\meter}) is shown here \citep{2014yCat.2328....0C}.}
%% No \tablerefs indicated
\end{deluxetable*}

For each target, the exposure time was set to reach at least $S/N \sim 100$. The target was kept centered on the slit without nodding the telescope. An A0 standard star was observed for telluric correction with $S/N \sim 150$ or greater. Flats and arc lamp exposures were taken for every object before slewing the telescope to remedy a known fringing issue in the flat fields of iSHELL. Dark frames were collected at the end of each night. The data were reduced using the IDL-based data reduction pipeline Spextool v5.0.3 \citep{2004paspspextool, 2003PASP..115..389V}. The data plotted in this paper are binned by three pixels to improve the $S/N$ and improve our ability to visually discern the line structure, but we use the original data to conduct our analyses. Because the slit of iSHELL is sampled by six pixels, binning by three pixels is Nyquist sampling the slit, so information about the line structure is preserved.

We make a few comments on two objects.

\textit{V1057 Cyg.} This is one of the original FUors described by \cite{1977ApJ...217..693H} and is thus an important member of the class. It is an exceptional object that has been shown to exhibit more drastic spectroscopic and photometric variability as it approaches quiescence \citep{10.1093/mnras/stt963, 2021ApJ...917...80S}. We initially observed this object in 2017, and then re-observed it in 2024 due to the atypical nature of its spectrum. We found significant changes to its K-band line structure along with a number of unusual spectral features in J- and M-band. V1057 Cyg is thus excluded from this paper because we cannot apply our line profile analysis to its spectrum.

\textit{PR Ori B.} This object was identified as a FUor relatively recently, following two significant brightening events of about 4 magnitudes and also owing to the fact that its near-infrared $R\sim2,000$ spectrum resembles that of other FUors \citep{2024ATel16776....1P}. We find that its CO lines exhibit an unusually high $v \sin i$ compared to other FUors in our sample. We therefore separate the data for PR Ori B by providing them in Appendix~\ref{sec:prori}.

\section{Methodology} \label{sec:methodology}

\begin{figure*}
    \centering
    \includegraphics[width=0.8\linewidth]{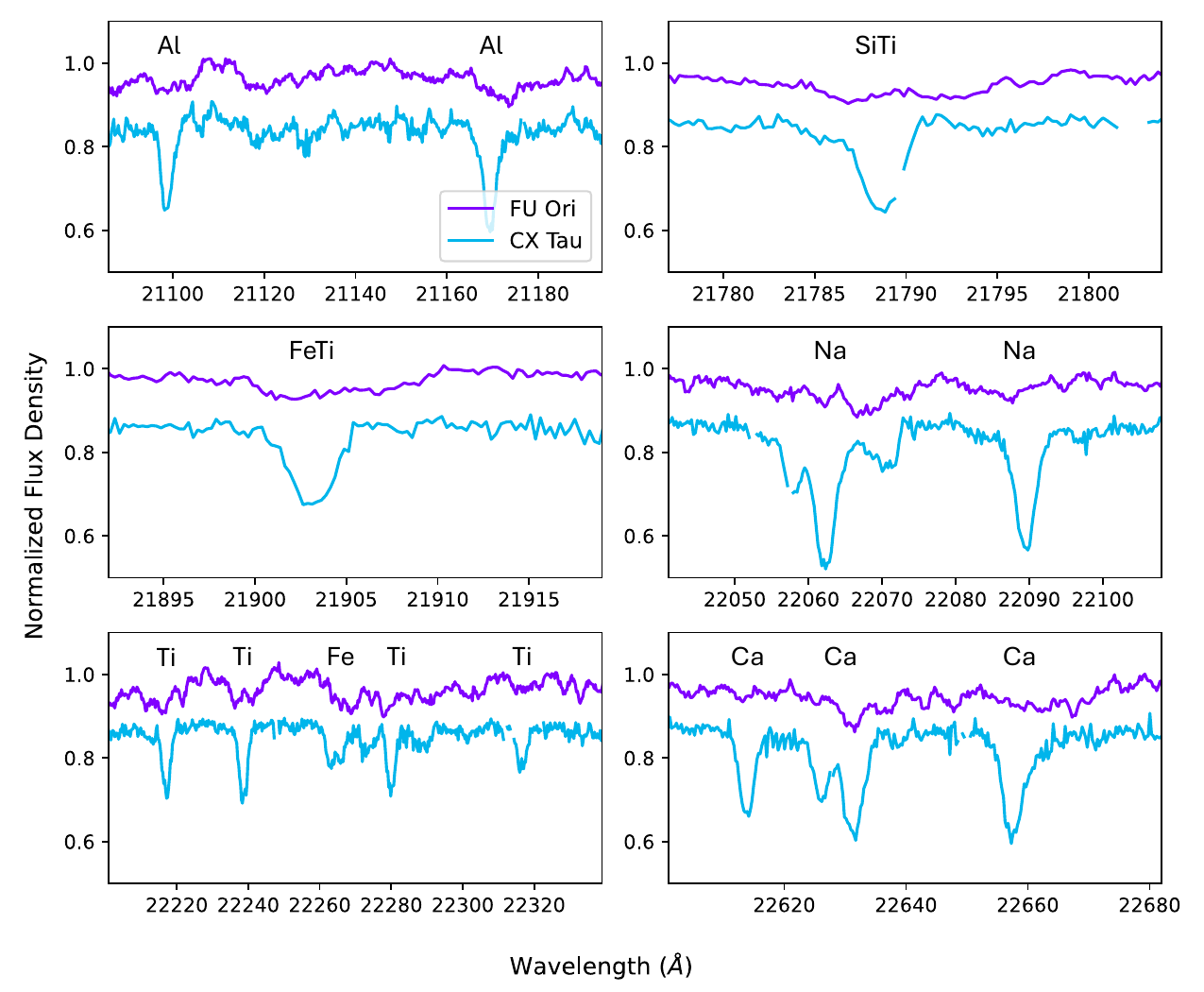}
    \caption{Regions of atomic absorption in the K-band spectrum of FU Ori compared to CX Tau, a typical Class II star (iSHELL data for CX Tau are from \cite{2022ApJ...925...21F}). These metal lines are typically useful as they are not impacted by extraneous material outside of the photosphere, unlike CO, but are not clearly identifiable in our FUor spectra.}
    \label{fig:atomic}
\end{figure*}

The observed spectrum is the intrinsic (unbroadened) spectrum convolved by the line profile. We fit the observed data by convolving template spectra with different types of line profiles over the first CO (2-0) overtone in K-band (22,900-\qty{23,215}{\angstrom}) and, when available, the aluminum lines in J-band (13,115-\qty{13,170}{\angstrom}). CO is thought to be impacted by outflow absorption in FUors \citep{2024ApJ...971...44C}, meaning that it may not be ideal for tracing the kinematics of the emitting photosphere. However, we still choose to use CO in our analysis because it is the only clearly distinguished absorption feature available in K-band (Fig.~\ref{fig:atomic}). We discuss the impact of a potential disk wind on our measurements later in Sec.~\ref{sec:discussion}.

\subsection{Template spectra} \label{sec:template}
\begin{figure}[h]
    \centering
    \includegraphics[width=\linewidth]{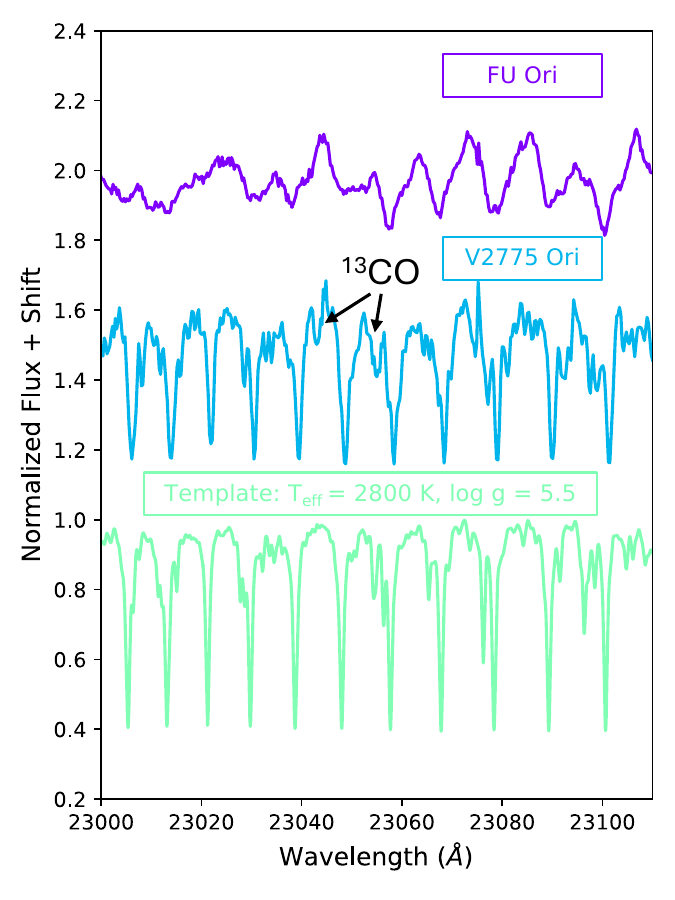}
    \caption{K-band CO lines of FU Ori, V2775 Ori, and a cool atmosphere model spectrum convolved with the instrument profile of iSHELL. The spectrum of FU Ori is multiplied by 1.5 to better view its spectral features. The spectra appear to be similar, but some objects including V2775 Ori show a small amount of \ce{^{13}CO} absorption over the model spectra.}
    \label{fig:vs_browndwarf_k2}
\end{figure}

\begin{figure}[h]
    \centering
    \includegraphics[width=\linewidth]{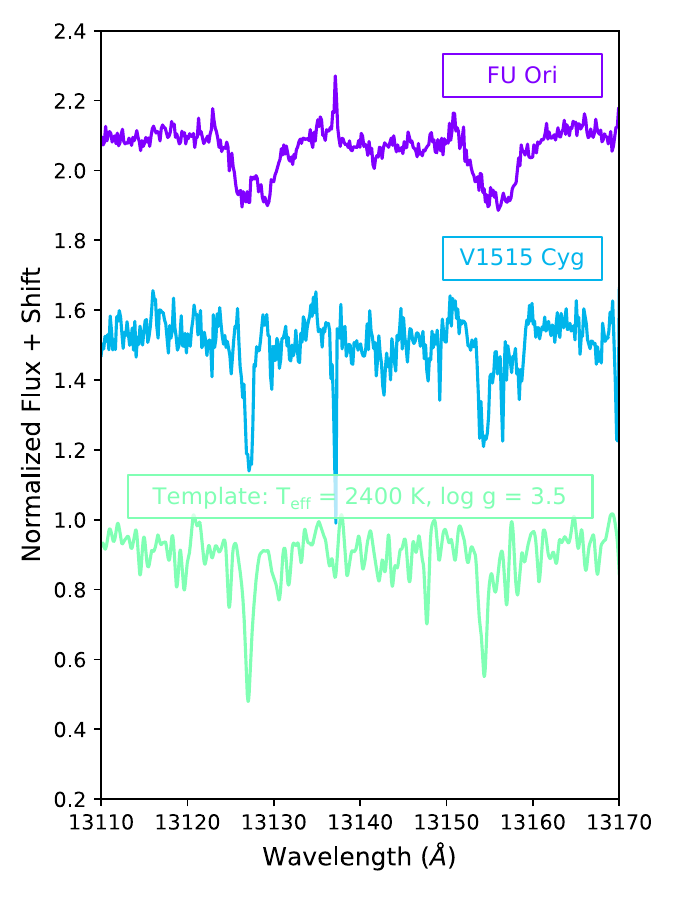}
    \caption{Same as Fig.~\ref{fig:vs_browndwarf_k2} except with Al lines in FU Ori and V1515 Cyg. FU Ori's spectrum is multiplied by 1.5. The template spectrum appears to be a reasonable match.}
    \label{fig:vs_browndwarf_j3}
\end{figure}

To approximate the intrinsic spectrum of our objects in J- and K-band, we employ the BT-Settl (CIFIST) model cool atmosphere spectra \citep{allard2013btsettl} for the reason that FUors viewed at $R\sim$ 1,200 with IRTF/SpeX appear strikingly similar to brown dwarfs \citep{connelley2018fuor}. Comparing template spectra to our targets which have narrow lines, shown in Figs.~\ref{fig:vs_browndwarf_k2} and \ref{fig:vs_browndwarf_j3}, we do not see large differences in the spectra that would drastically impact the fitted line profile. There may be slight differences in molecular abundances that can change the fitted line profiles.

\begin{figure}[htpb]
    \centering
    \includegraphics[width=\linewidth]{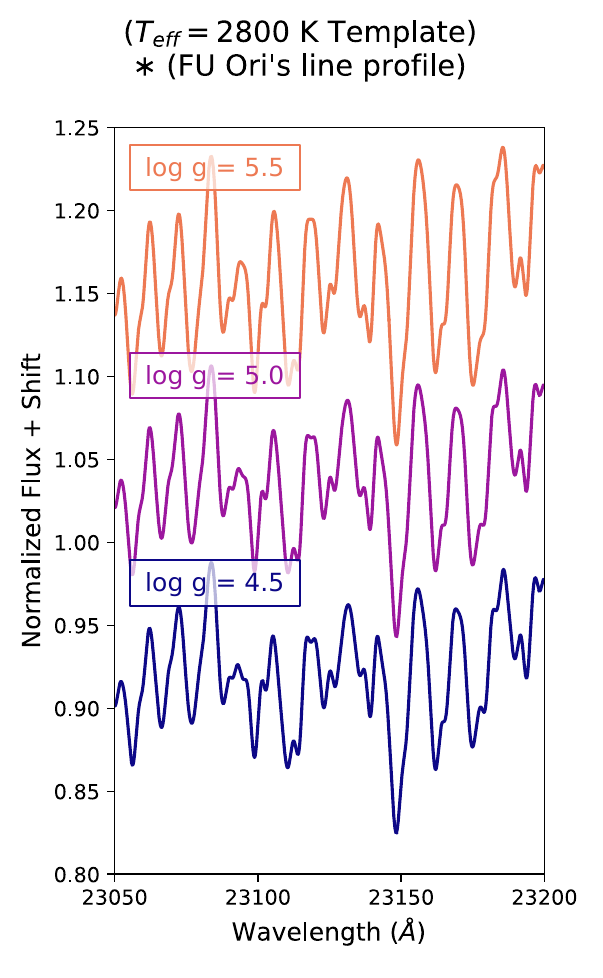}
    \caption{Template spectra with different surface gravities convolved with the best-fit asymmetrical, double-peaked line profile of FU Ori (shown later in Fig.~\ref{fig:corner_fuori-2gauss}). The surface gravity noticeably impacts the shapes and depths of the lines.}
    \label{fig:logg_impact}
\end{figure}

We use a range of templates spanning from $T_{eff}=$2000 to \qty{3400}{\kelvin} and $\log g = 4.5$ to 5.5 dex. Although the templates with larger surface gravity have broader lines, the wings of the lines are also broadened in a particular manner so that there is not necessarily degeneracy between surface gravity and the $v \sin i$ computed from our line profiles. Figure~\ref{fig:logg_impact} shows that an increase in $\log g$ tends to make the lights slightly smoother and steeper for a fixed $T_{eff}$, meaning that it is distinguished from a pure rotational broadening.

\begin{figure*}[t]
    \centering
    \includegraphics[width=0.9\linewidth]{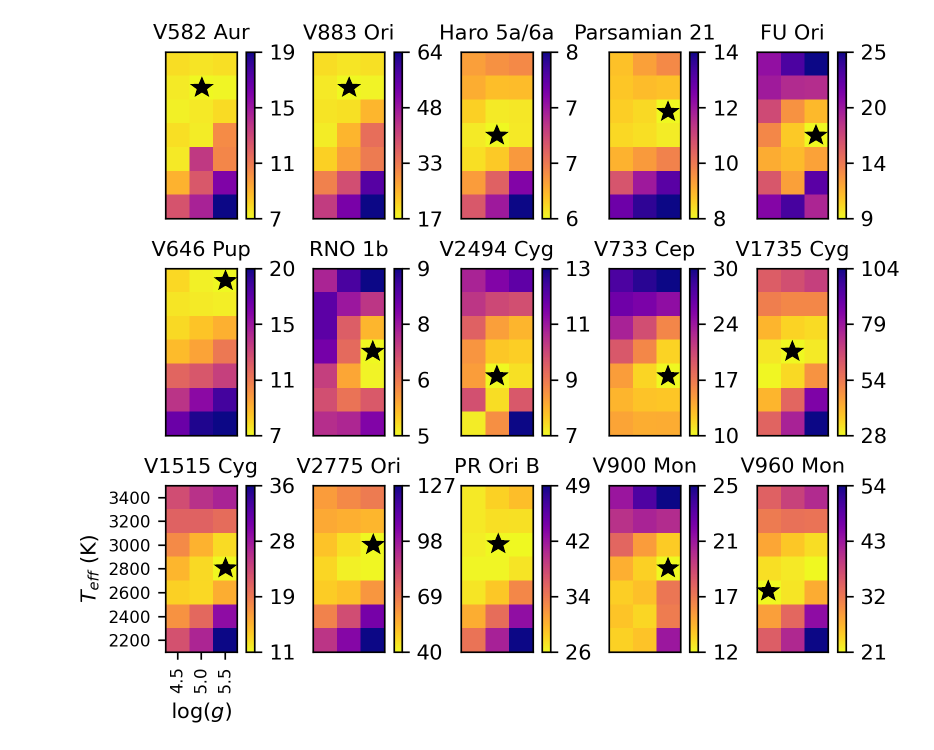}
    \caption{Reduced chi-squared ($\chi^{2}_{\nu}$) values for each object fitted over a grid of template spectra in K-band, with decreasing $\chi^{2}_{\nu}$ corresponding to brighter colors. The best template spectra are marked with stars. This figure demonstrates that different objects benefit from using different templates. There appears to be some degeneracy between $T_{eff}$ and $\log g$. However, we combine the results of all template spectra by averaging over the line profiles, weighting by the $\chi^2$ values, so the measured line profiles are not impacted by this.}
    \label{fig:chi2}
\end{figure*}

We apply a grid of template spectra with different $T_{eff}$ and $\log g$ to each object. Figure~\ref{fig:chi2} shows that it is beneficial to describe our objects with a variety of template spectra, rather than using a single template for all objects. The surface gravities of the best template spectra for our objects appear to be quite large. We are not properly fitting stellar parameters to many photospheric lines, so the $T_{eff}$ and $\log g$ of the best template spectra cannot really be used to form a robust interpretation of the physical properties of our objects. This is especially true when considering the CO lines, which are known to be affected by accretion, circumstellar absorption, and other factors in young stars. It is therefore difficult for us to interpret these values.

For each template spectrum, we fit line profiles by using the Markov Chain Monte Carlo (MCMC) technique implemented by the package Python package emcee \citep{foreman2013emcee}. The fitted parameters are given by the median of the marginal posterior distributions and the uncertainties are determined by the 16th and 84th percentiles. We describe how each type of line profile is parameterized in Sec.~\ref{sec:lp}.

Each template spectrum results in a slightly different line profile. To combine the results of all the template spectra for each object, we use a process called Bayesian model averaging (BMA), which essentially averages the models after they have been weighted by their respective likelihoods ($\chi^2$ value) and priors \citep{hoeting1999bayesian, hinne2020conceptual}. The standard method for BMA described by \cite{hoeting1999bayesian} is
\begin{align}
    \text{E}(p|D)& = \sum_{k=0}^{K}\hat{p}_{k} \text{P}(M_k|D) \\
    \text{Var}(p|D) &= \sum_{k=0}^{k} (\text{Var}(p|D,M_k) +\hat{p}_k^2) \text{P}(M_k|D)  - \text{E}(p|D)^2
\end{align}
where $\text{E}(p|D)$ are the expected (combined) parameters given the data $D$ and $\text{Var}(p|D)$ are the associated variances (errors). $M_k$ is the $k$th template spectrum and $P(M|D)$ is the probability of the model given the data, which can be computed from the $\chi^2$ value and prior. $\hat{p}_k = \text{E}(p|D,M_k)$ are the parameters fitted from each template. We find that in practice there is not much difference in the combined parameters and uncertainties compared to those produced by the best template by itself.

\subsection{Line profile fitting} \label{sec:lp}
We use a linear combination of Gaussians to model the line profile for each object, allowing the relative amplitudes, centers, and widths of each Gaussian to vary. We begin by fitting every object with a single Gaussian as well as a combination of two Gaussians. Then, we compute the F-statistic from the reduced chi-squared ($\chi^2_v$) values to determine whether the two Gaussian model provides a significantly better fit (p-value $>$ 0.95). If not, the single Gaussian is kept. Otherwise, one more Gaussian is added to the model and the F-test is performed again with the intent repeating the process until a sufficient number of Gaussians is found. In actuality, we find that all of the objects are well-described by two or less peaks.

A combination of Gaussians is unlikely to be the exact functional form of the line profiles. Our intention is to gain a cursory estimate of the linewidth and the presence of multiple peaks rather than creating a physically robust model. For objects that exhibited double-peaked line profiles, we also fit a Keplerian disk profile.

Prior to fitting each spectrum, we normalize the continuum by using the \verb|specnorm| Python script.\footnote{\url{https://python4esac.github.io/plotting/specnorm.html}} This removes the slope of the continuum. It is often difficult to determine the exact continuum flux level due to the proliferation of molecular absorption features that contaminate the continuum, especially at K-band. To ameliorate this, we fit the parameter $\delta y$ which allows the model spectrum to translate along the y-axis by a miniscule amount.

\subsection{Implementation of line profiles}
We always begin by convolving each line profile with the instrument profile of iSHELL measured by \cite{flores2019measuring}. This is equivalent to convolving the the model spectrum with the instrument profile because convolution is associative. 

\begin{figure*}[h]
    \centering
    \includegraphics[width=\linewidth]{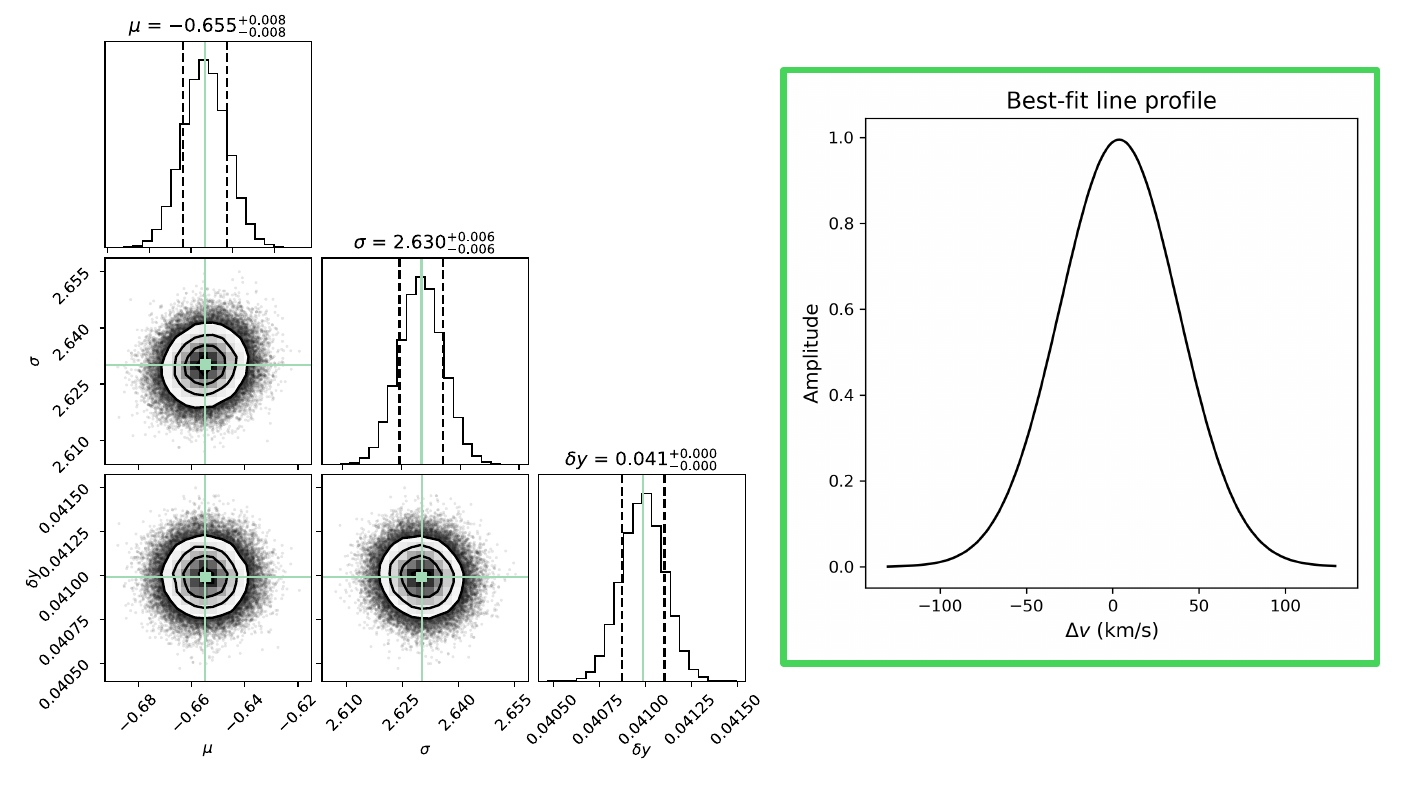}
    \caption{Posterior distribution for the single Gaussian convolution kernel for FU Ori. This figure was generated using the corner package \citep{corner}. The best-fit line profile is shown in the panel to the right, outlined in green.}
    \label{fig:corner_fuori-1gauss}
\end{figure*}

A Gaussian line profile is described by
\begin{equation}
    \Phi(\Delta \lambda) = e^{-\frac{1}{2}(\frac{\Delta \lambda - \mu}{\sigma})^2}
\end{equation}
where $\mu$ is the mean and $\sigma$ is the standard deviation. With a single Gaussian, there is no need for an amplitude term because the convolution kernel is normalized. We fit the parameters $\mu,\sigma$, and the previously described $\delta y$. The majority of the Doppler shift is eliminated by cross-correlating the spectrum with a broadened template, but the residual shift is removed by allowing the mean to vary. An example posterior distribution is shown in Fig.~\ref{fig:corner_fuori-1gauss}. Also note that the equations are written in terms of wavelength for convenience. The parameters are converted to velocity space later by using the central wavelength of fitted section of the spectrum.

\begin{figure*}[htpb]
    \centering
    \includegraphics[width=\linewidth]{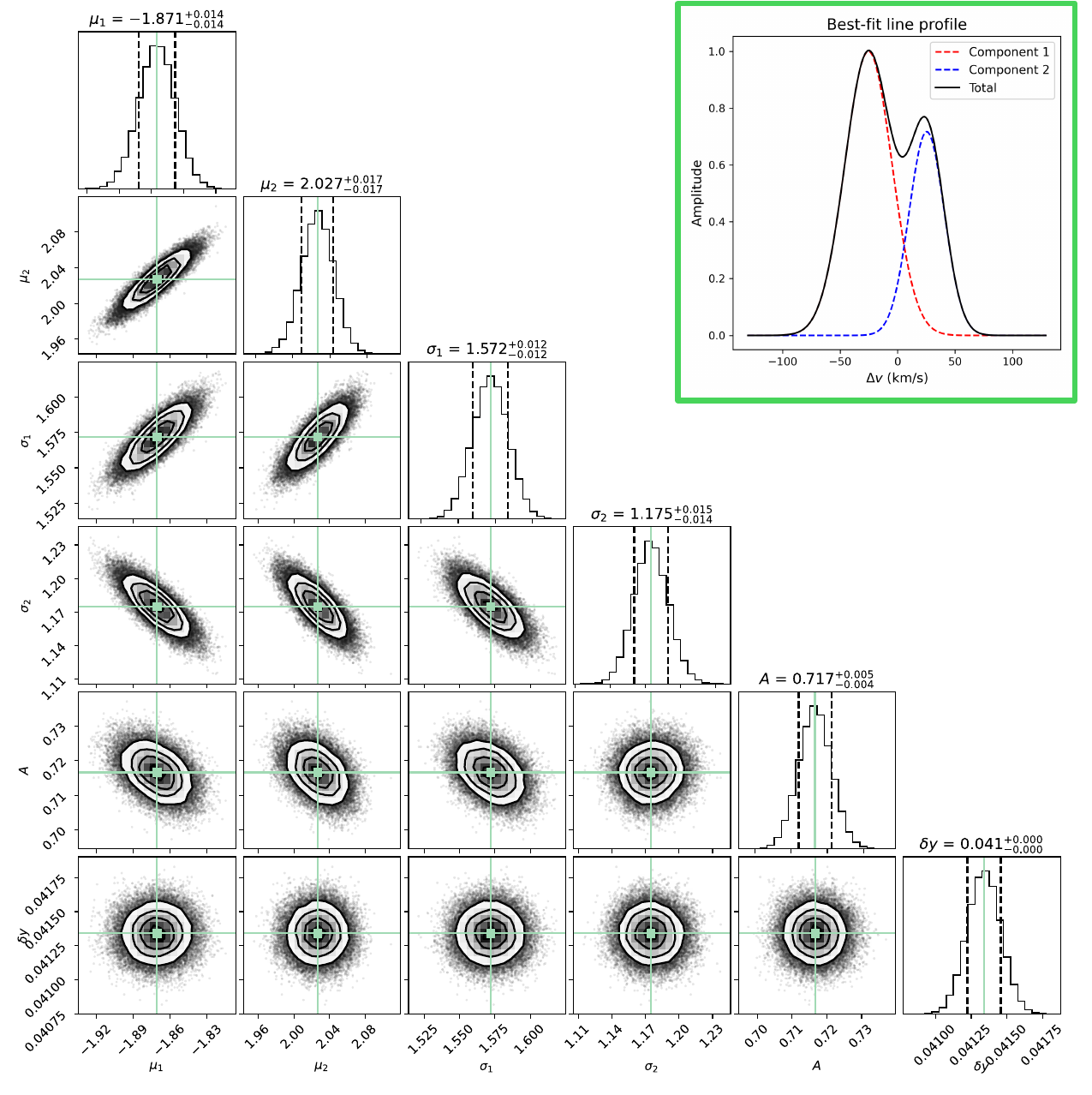}
    \caption{Sampled posterior distribution for the best-fit double Gaussian convolution kernel for FU Ori. The two gaussian components of the line profile are shown as colored, dotted lines.}
    \label{fig:corner_fuori-2gauss}
\end{figure*}

For two Gaussians, the line profile is
\begin{equation}
     \Phi(\Delta \lambda) = e^{-\frac{1}{2}(\frac{\Delta \lambda - \mu_1}{\sigma_1})^2} + A e^{-\frac{1}{2}(\frac{\Delta \lambda - \mu_2}{\sigma_2})^2}
\end{equation}
where $A$ is the relative amplitude. To minimize degeneracy, we enforce $\mu_1 \leq \mu_2$ and also use Gaussian priors on all parameters except for $A$ and $\delta y$. The priors on $\mu_1$ and $\mu_2$ discourage a ``phase shift" in the CO lines at K-band, while the priors on $\sigma_1$ and $\sigma_2$ prevent the widths of the Gaussians from going to infinity. The Gaussian priors have a standard deviation of five and a mean of one. They encapsulate a much larger region of parameter space than the sampled posterior distribution (see Fig.~\ref{fig:corner_fuori-2gauss}).

\begin{figure*}[htpb]
    \centering
    \includegraphics[width=\linewidth]{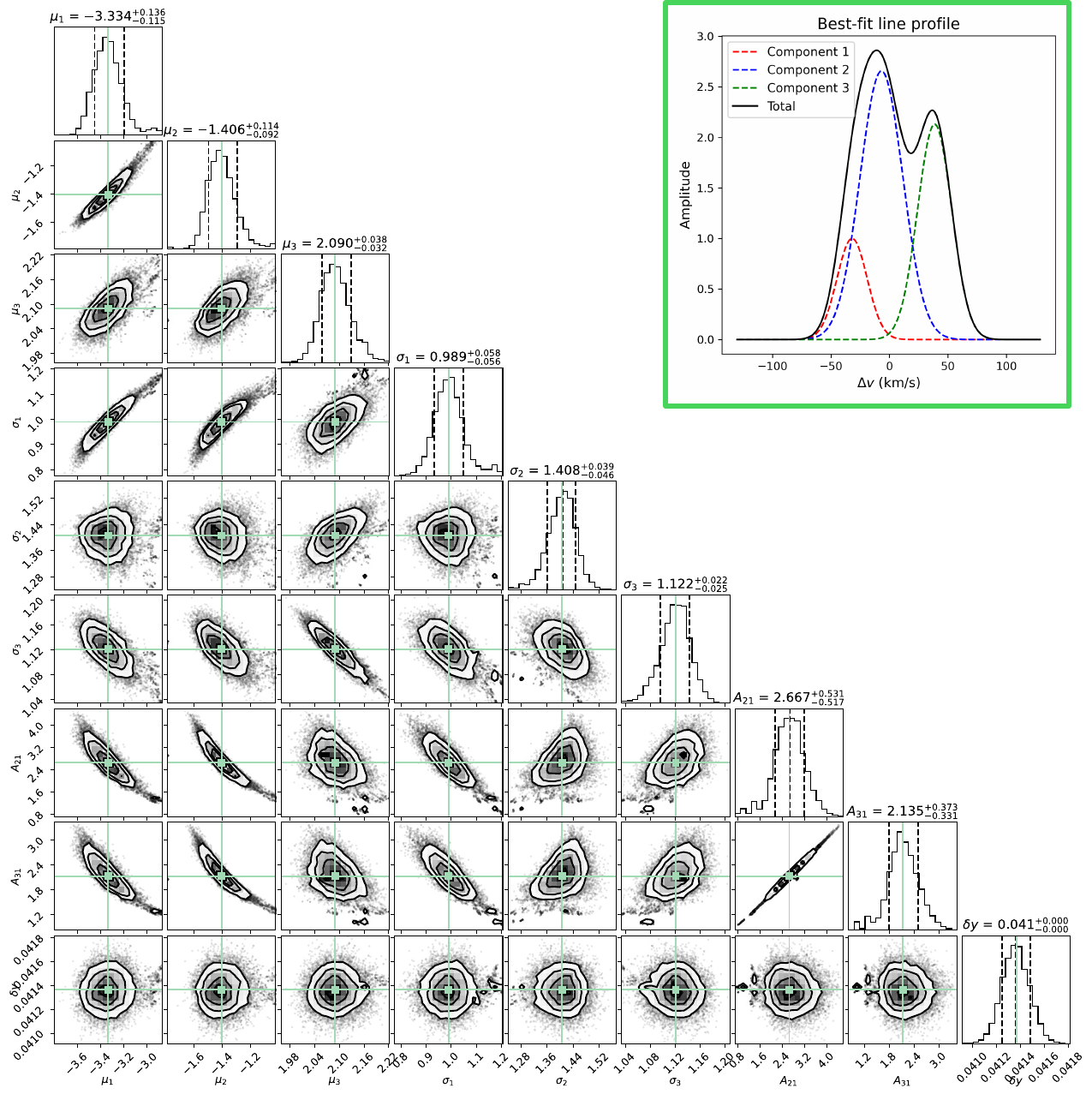}
    \caption{Sampled posterior distribution for the best-fit triple Gaussian convolution kernel for FU Ori. The three gaussian components of the line profile are shown as colored, dotted lines.}
    \label{fig:corner_fuori-3gauss}
\end{figure*}

The line profile for three Gaussians is implemented as
\begin{multline}
     \Phi(\Delta \lambda) = e^{-\frac{1}{2}(\frac{\Delta \lambda - \mu_1}{\sigma_1})^2} + A_{21} e^{-\frac{1}{2}(\frac{\Delta \lambda - \mu_2}{\sigma_2})^2} \\
     + A_{31} e^{-\frac{1}{2}(\frac{\Delta \lambda - \mu_3}{\sigma_3})^2}
\end{multline}
where the $A_{21}$ and $A_{31}$ are the amplitudes relative to the first Gaussian. Similar priors are applied as in the two Gaussian case to avoid degeneracy. Figure~\ref{fig:corner_fuori-3gauss} shows that all parameters converge.

\begin{figure*}[htpb]
    \centering
    \includegraphics[width=\linewidth]{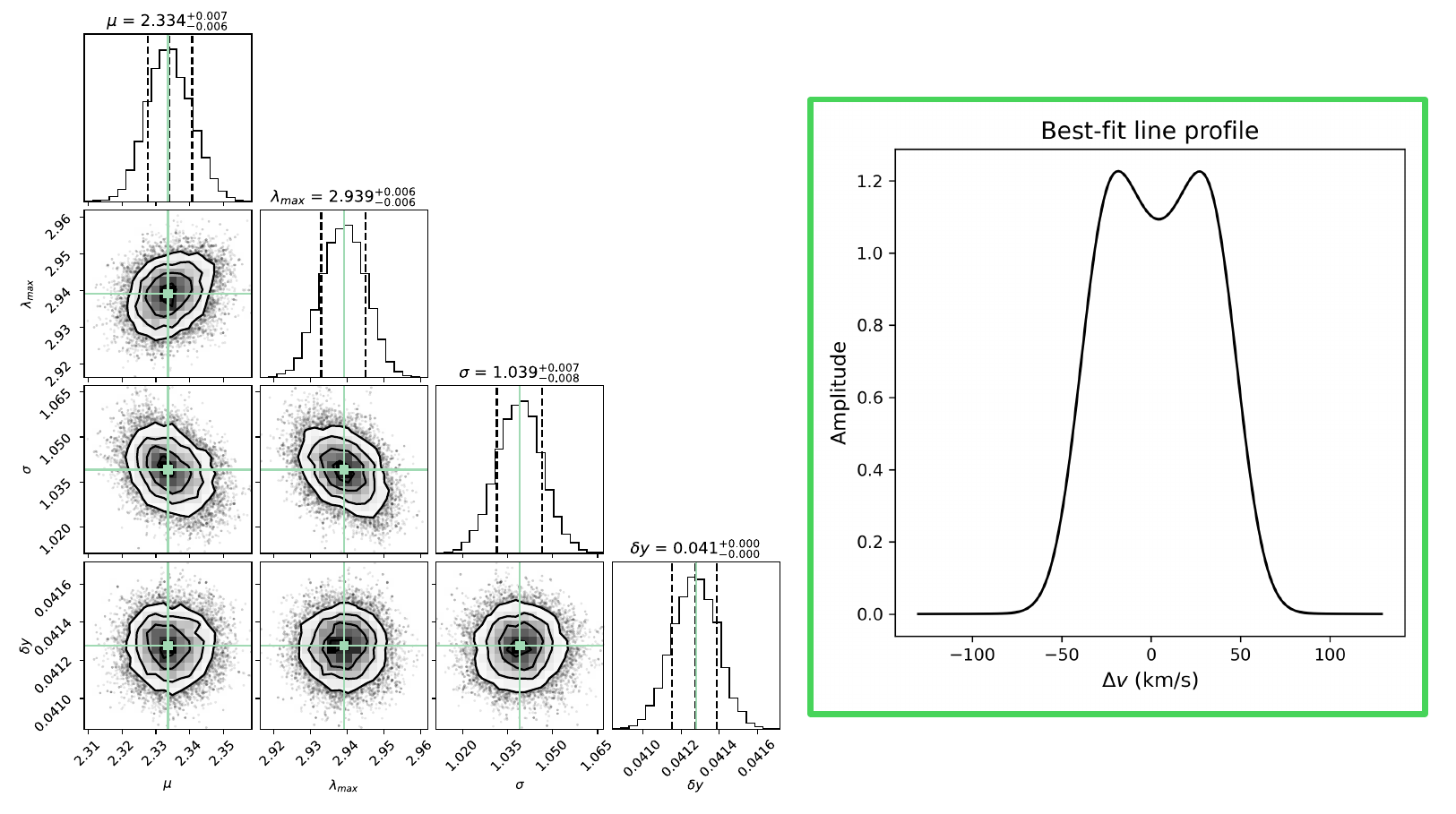}
    \caption{Sampled posterior distribution for the best-fit Keplerian convolution kernel for FU Ori.}
    \label{fig:corner_fuori-kep}
\end{figure*}

We implement the ``double-U" Keplerian line profile as
\begin{equation}
    \Phi'(\Delta \lambda) = \begin{cases} 
        \Big[ 1 - \Big( \frac{\Delta \lambda}{\lambda_{max}} \Big)^2 \Big]^{-1/2} & \text{if } |\Delta \lambda| < \lambda_{max} \\
        0 & \text{if } |\Delta \lambda| \geq \lambda_{max}
        \end{cases} \\
\end{equation}
\begin{equation}
    \Phi(\Delta \lambda) = \Phi'(\Delta \lambda)*G(\Delta \lambda)
\end{equation}
where $\Phi'(\Delta \lambda)$ is the Keplerian line profile described in \cite{hartmann1985nature} and $G(\Delta \lambda)$ is a Gaussian with a standard deviation of $\sigma$ and mean $\mu$ that smooths out this function. The Gaussian smoothing is usually attributed to turbulence driven by thermal motions and magnetic fields in the inner disk \citep{lodato2003probing, 2004ApJ...609..906H, 2020MNRAS.495.3494Z}, and has been found to help reproduce high-resolution optical spectra of FUors assuming an accretion disk model \citep{2023ApJ...958..140C}. Figure~\ref{fig:corner_fuori-kep} demonstrates that our implementation of the Keplerian line profile is well-behaved. 

\subsection{M-band spectra}

\begin{figure*}[htpb]
    \centering
    \includegraphics[width=0.9\linewidth]{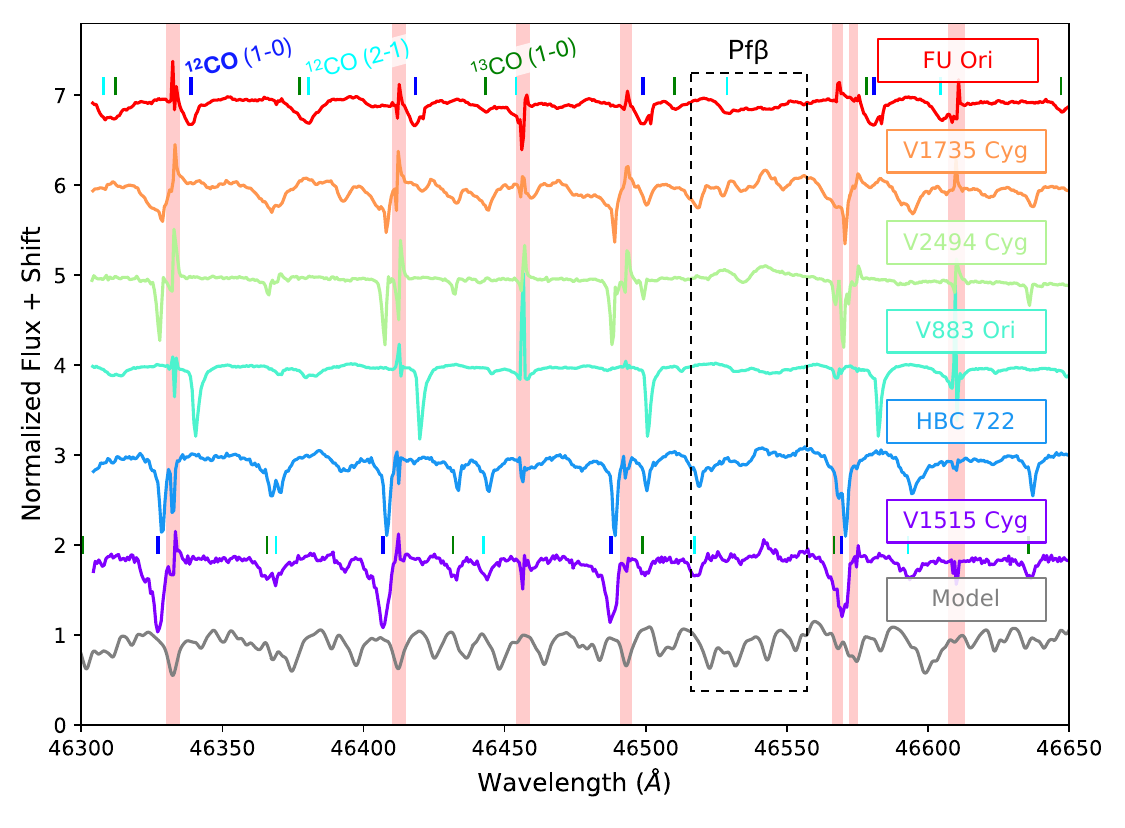}
    \caption{A selected portion of the observed M-band spectra. Colored dashes above the spectra of FU Ori and V1515 Cyg identify different CO lines. Regions of strong telluric contamination are shaded in red. The model spectrum (example shown in gray: $T_{eff}=\qty{2800}{\kelvin}, \log g=5.0$) are not a good match to the observed data. The profiles of the CO lines appear markedly different from the K-band data. For example, V883 Ori is one of the fastest rotating objects in our sample but has narrower CO lines than V1515 Cyg, which shows very little line broadening in K-band. Also striking is the depth of the CO lines, which appear nearly saturated for some objects. The approximate location of Pf$\beta$ is indicated; in emission, it is used as an indicator for magnetospheric accretion but is weak or absent in several of our objects.}
    \label{fig:co_m2}
\end{figure*}

FUors may appear spectroscopically similar to brown dwarfs in J- and K-band, but we find that this is not true in M-band. A cursory glance at the spectra in Fig.~\ref{fig:co_m2} reveals a smoother continuum than the cool atmosphere model, perhaps due to increased thermal emission from the dust. As we will show in the following section, the profiles of the CO lines seen here also appear to lack a correlation to those in J- and K-band.

All of this is to say that we cannot model the M-band data by convolving the template spectra. Instead, we average ten $^{12}$CO (1-0) lines between 4.63 to \qty{4.8}{\micro\meter}--chosen for being mostly uncontaminated by telluric absorption and other features--and fit a linear combination Gaussians. To identify the lines, we used the spectools\_ir package to query the HITRAN molecular absorption database \citep{salyk2022csalyk, 2026JQSRT.35309807G}. A basic least-squares fit is performed using the \verb|curve_fit()| function provided by the Scipy package \citep{2020SciPy-NMeth}. Directly fitting Gaussians to the lines is fundamentally different from convolution as only the latter preserves the equivalent width. The $v \sin i$ measurements in M-band should be taken as overestimates as the lines appear to near saturation, which can also increase the linewidth.

\section{Results} \label{sec:results}
\begin{figure*}[p!]
    \centering
    \includegraphics[width=\linewidth]{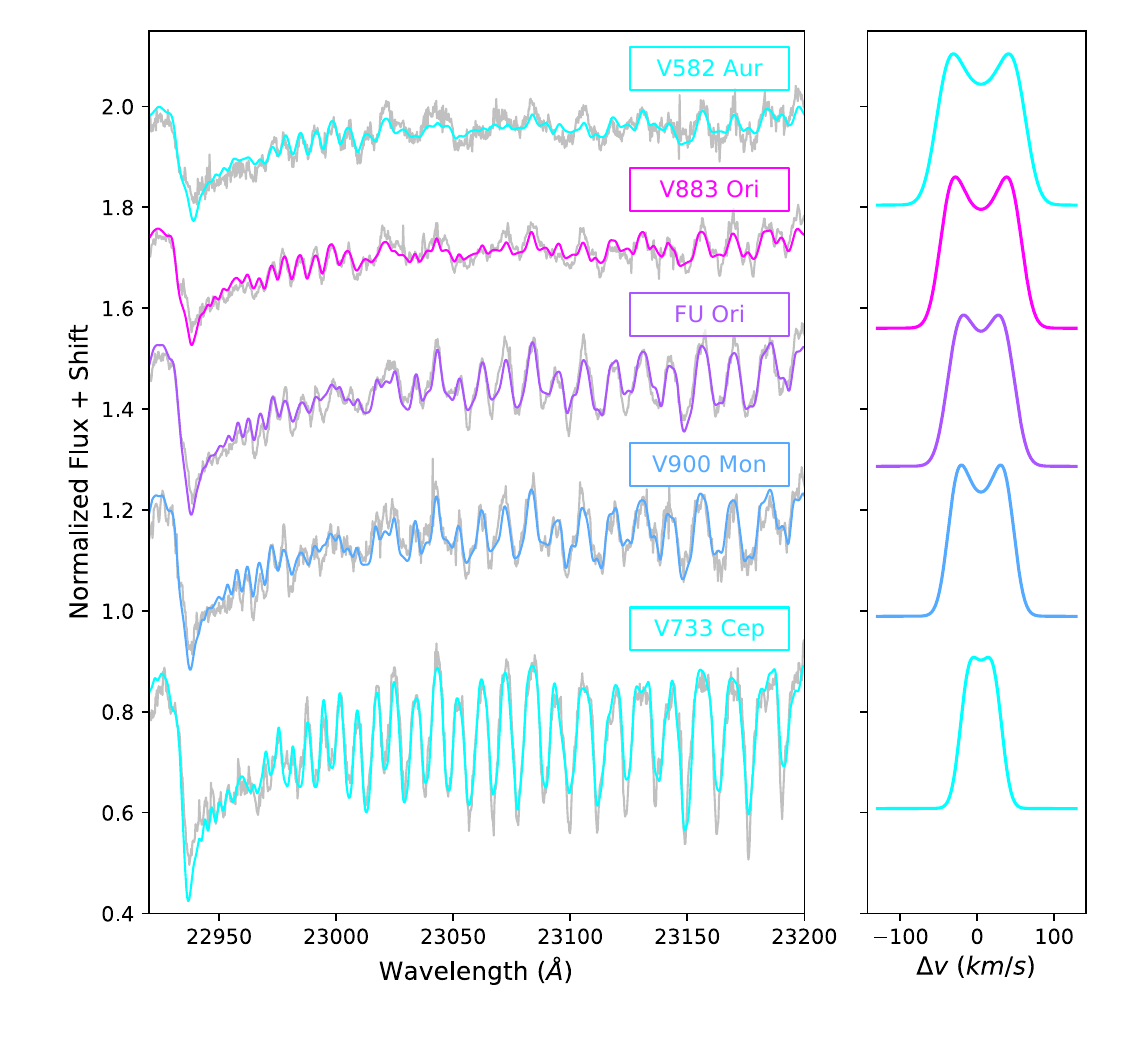}
    \caption{Keplerian line profiles fitted to the objects with double-peaked line profiles. The quality of the fit appears to be similar as for two Gaussians.}
    \label{fig:spectra_k2-kep}
\end{figure*}

\begin{figure*}[p!]
    \includegraphics[width=\linewidth]{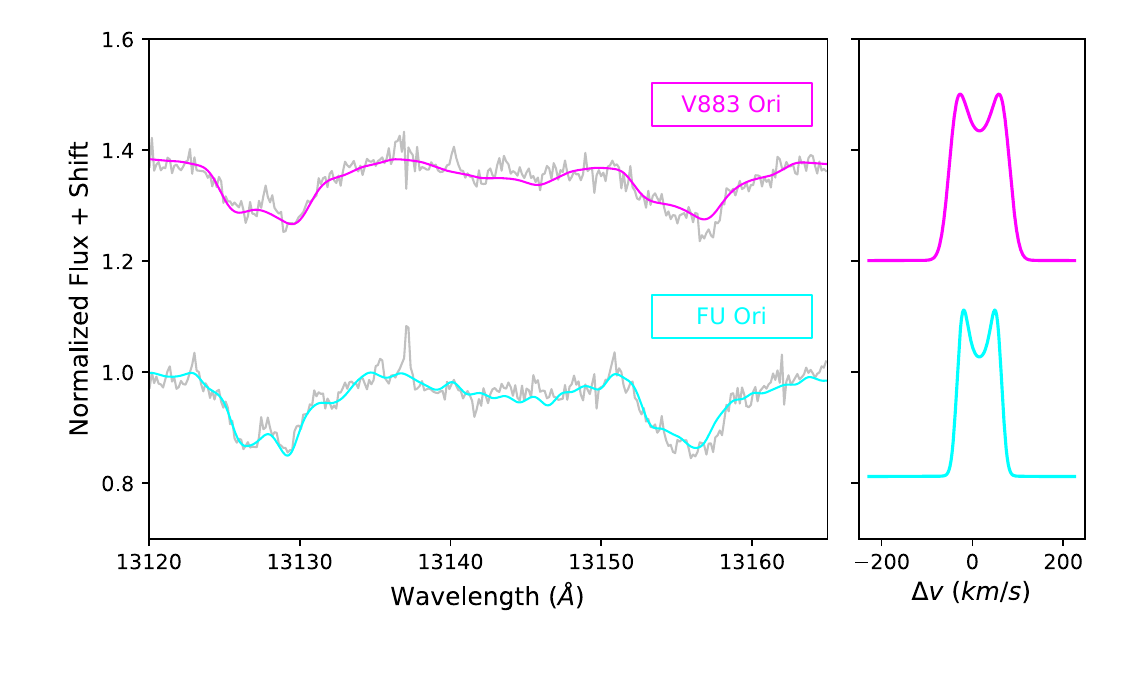}
    \caption{Same as Fig.~\ref{fig:spectra_k2-kep} but with the Al lines in J-band. Again, the fit appears to be good.}
    \label{fig:spectra_j3-kep}
\end{figure*}

Figures~\ref{fig:spectra_k2} and \ref{fig:spectra_j3-2} show the line profiles modeled as a linear combination of Gaussians in order of decreasing linewidth. Five of the objects are fitted by double-peaked line profiles, which would be expected for a Keplerian disk. Double-peaked line profiles appear to be more frequent in objects with larger $v\sin i$. Figures~\ref{fig:spectra_k2-kep} and \ref{fig:spectra_j3-kep} show Keplerian line profiles fitted to these objects. Determining whether the Keplerian or double Gaussian line profile is better in a statistical sense is not straightforward because there is no clearly defined null hypothesis, e.g., that additional parameters are zero (as in the case of the F-test for nested models). However, we can see that the Keplerian line profile provides a qualitatively similar fit. 

\begin{figure*}[p!]
\renewcommand{\theHfigure}{\arabic{figure}a}
    \centering
    \includegraphics[width=\linewidth]{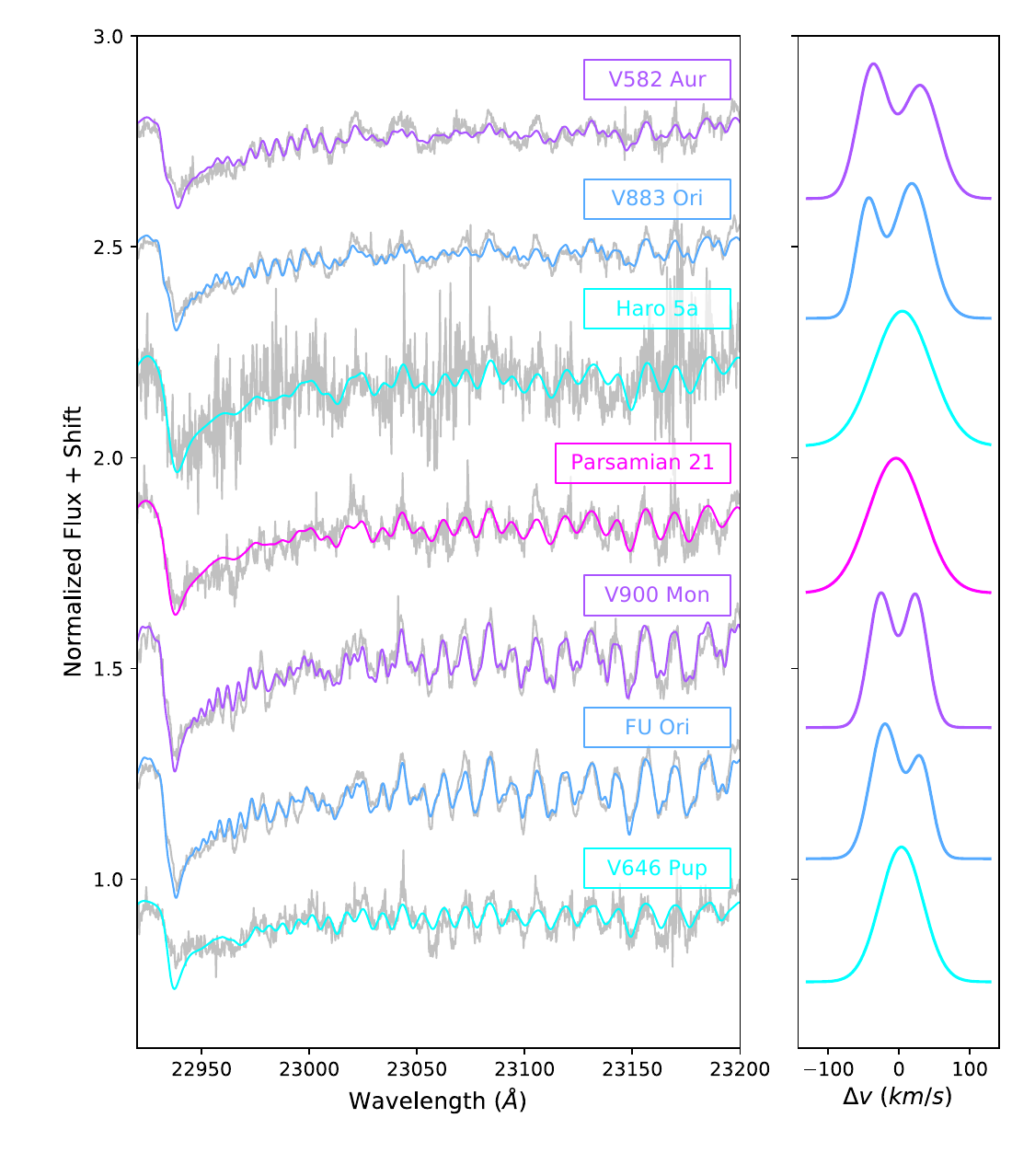}
    \caption{Line profile fits to the first CO overtone in K-band in order of decreasing linewidth. The line profiles here are described as a linear combination of Gaussians (right column). The observed spectra are shown in gray while the convolved templates are shown as colored lines. The poor fit to PR Ori B is discussed in Sec.~\ref{sec:results}.}
    \label{fig:spectra_k2}
\end{figure*}

\begin{figure*}[p!]
    \addtocounter{figure}{-1}
    \renewcommand{\theHfigure}{\arabic{figure}b}
    \centering
    \includegraphics[width=\linewidth]{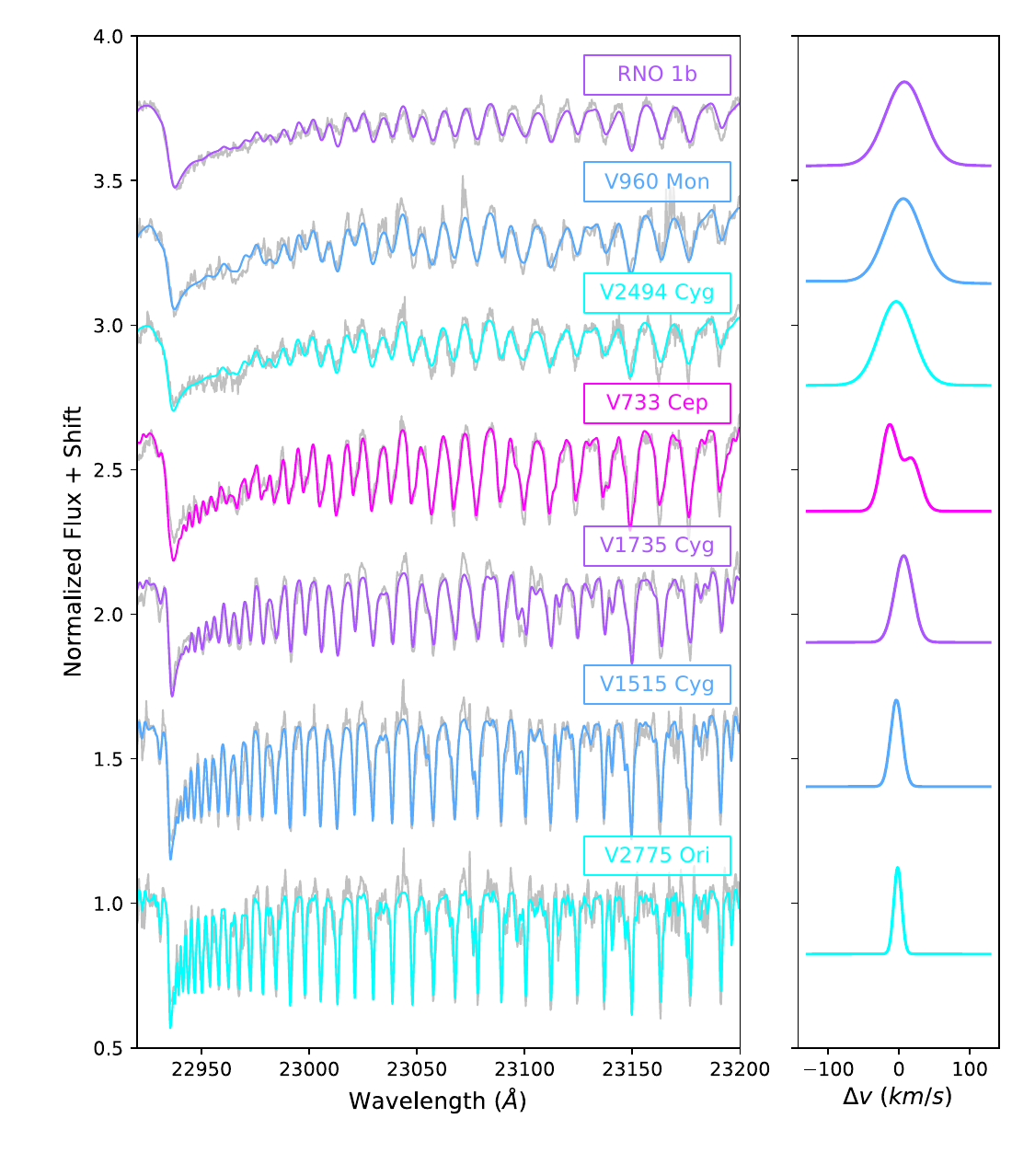}
    \caption{Continued.}
    \label{fig:spectra_k2-2}
\end{figure*}

\begin{figure*}[p!]
\renewcommand{\theHfigure}{\arabic{figure}a}
    \centering
    \includegraphics[width=\linewidth]{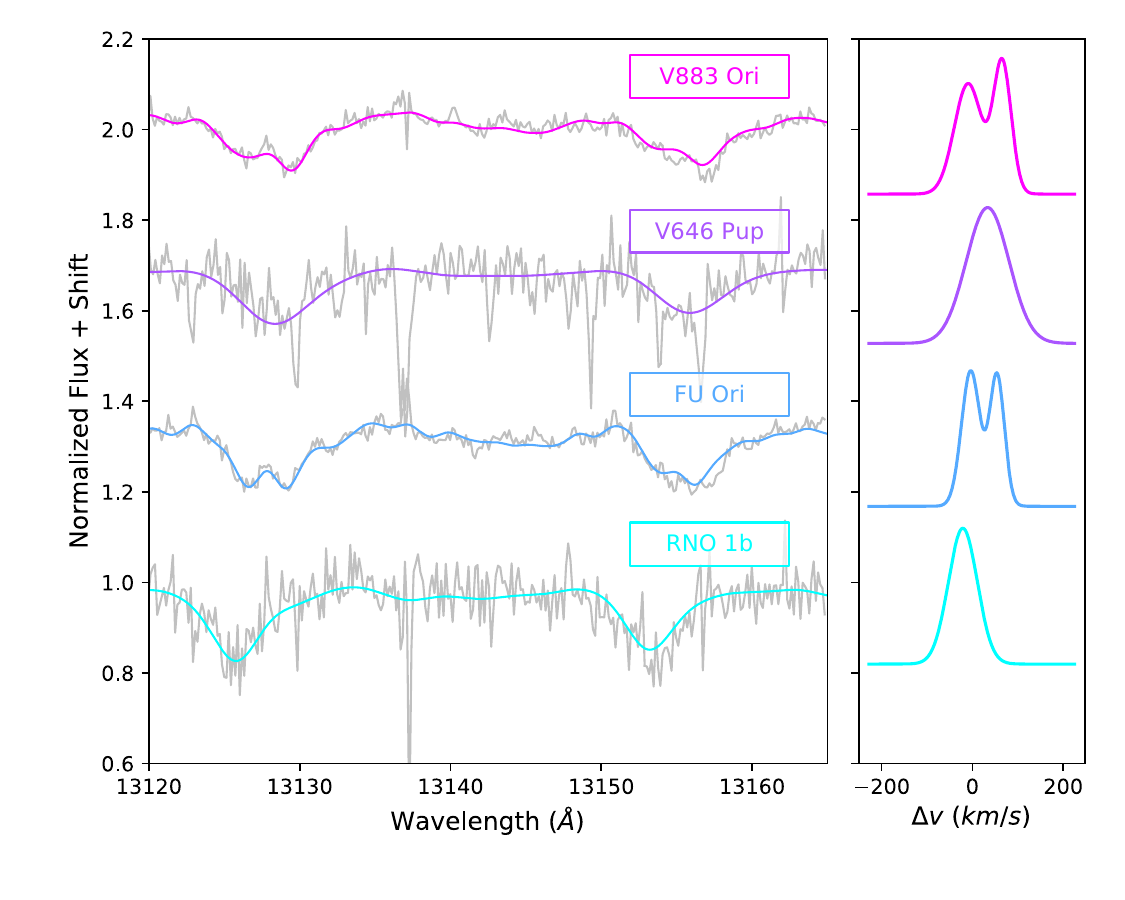}
    \caption{Same as Fig.~\ref{fig:spectra_k2}, except with the Al lines in J-band.}
    \label{fig:spectra_j3}
\end{figure*}

\begin{figure*}[p!]
    \addtocounter{figure}{-1}
    \renewcommand{\theHfigure}{\arabic{figure}b}
    \centering
    \includegraphics[width=\linewidth]{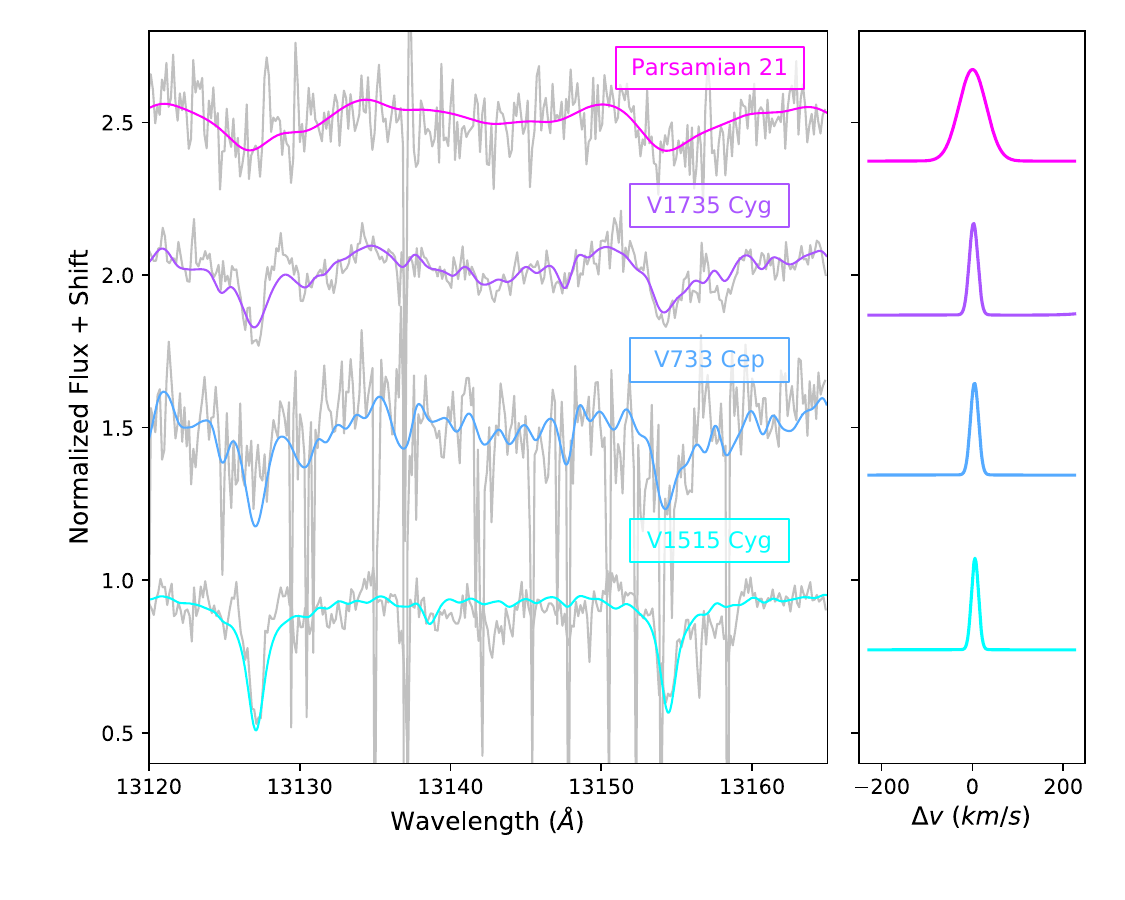}
    \caption{Continued.}
    \label{fig:spectra_j3-2}
\end{figure*}

\begin{figure}
    \centering
    \includegraphics[width=\linewidth]{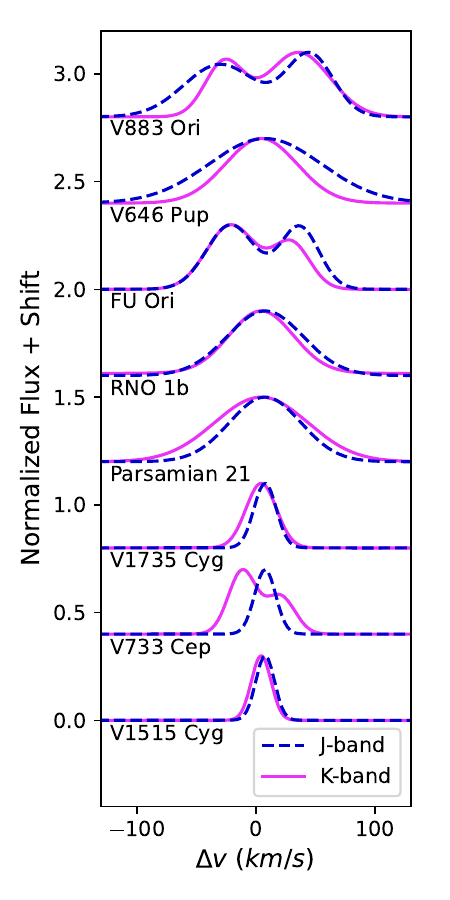}
    \caption{Comparison of the J-band (dashed blue) and K-band (pink) line profiles. They appear to be similar to each other save for V733 Cep, which has low SNR data in J-band.}
    \label{fig:lp-compare}
\end{figure}

\begin{figure}
    \centering
    \includegraphics[width=\linewidth]{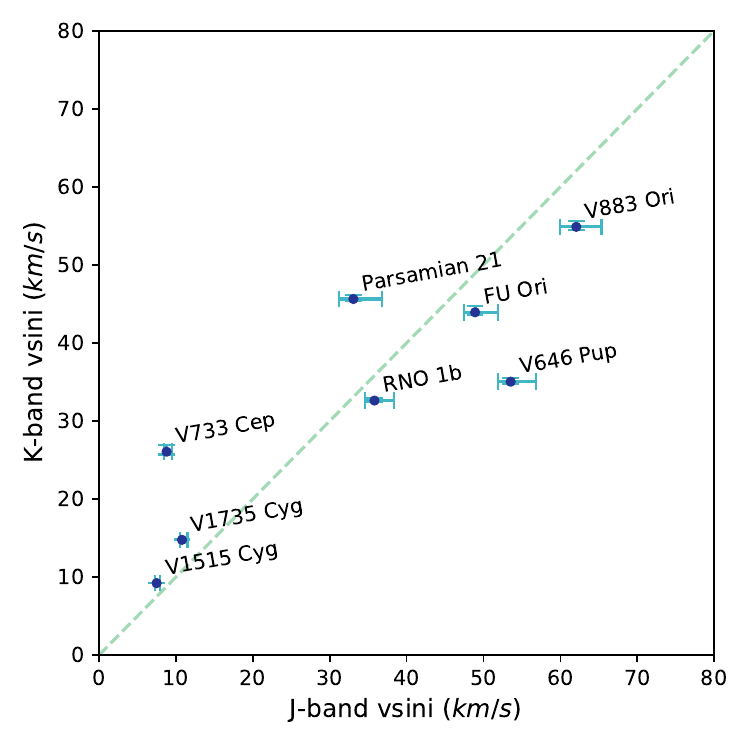}
    \caption{K- versus J-band $v\sin i$. The dotted line has a slope of 1. We do not clearly see the wavelength dependence of the linewidth predicted by a basic Keplerian disk model, which predicts that the objects will fall on a line with a slope less than 1. This does not necessarily mean that the accretion disk model is incorrect (see Sec.~\ref{sec:discussion}). PR Ori B is excluded from this plot since we could not obtain a good fit to its spectrum.
    }
    \label{fig:linewidths_JK}
\end{figure}

The J- and K-band line profiles generally appear to be consistent with each other (Fig.~\ref{fig:lp-compare}). The exception is V733 Cep which is only fit by a double-peaked profile in K-band. It is apparent that the J-band data has low SNR and that the linewidth is underestimated. A basic Keplerian disk model also predicts that there should be narrower linewidths at longer wavelengths because the cooler outer regions of the disk rotate more slowly. The measured $v \sin i$ are displayed in Table~\ref{tab:vsini}, which are the half-width at half-maximum of the line profiles converted to velocity space. The uncertainties are derived from the line profiles corresponding to the 16th and 84th percentiles of the posterior distribution. Figure~\ref{fig:linewidths_JK} shows that $v \sin i$ in J- and K-band are correlated, but there is not a clear decrease in linewidth with redder wavelengths. Optical $v\sin i$ are listed in Table~\ref{tab:vsini} when available in the literature. All of the retrieved optical $v\sin i$ are larger than what we measure in the near-infrared.

\begin{figure*}[p!]
    \centering
    \includegraphics{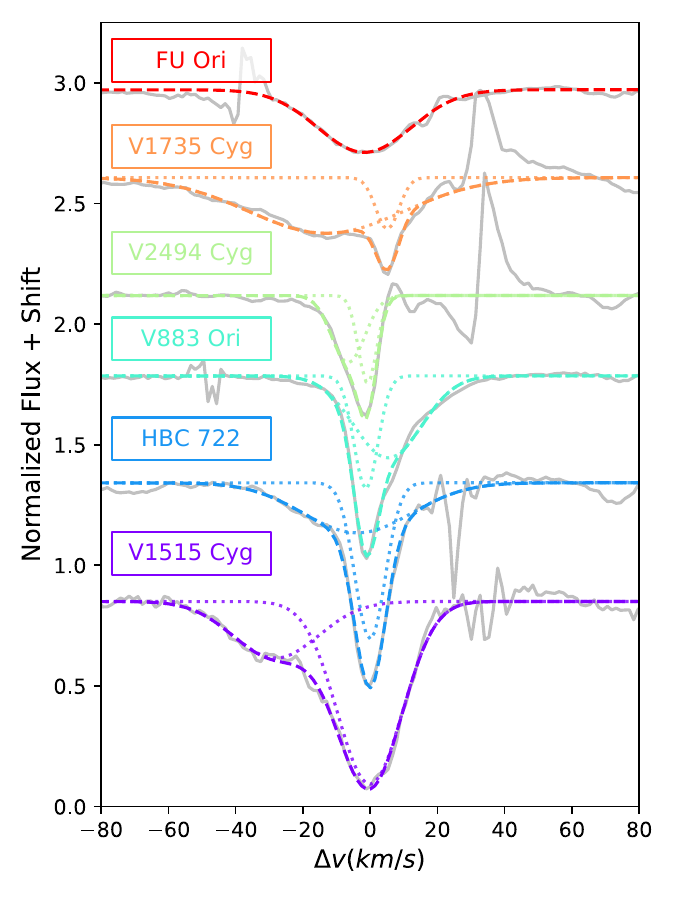}
    \caption{Stacked CO lines in M-band directly fitted with a sum of two Gaussians (dotted lines for the two components and dashed lines for the sum of the two). The lines here appear to be unrelated to what is seen in J- and K-band.}
    \label{fig:m2_linefit}
\end{figure*}

On the other hand, the M-band line profiles (Fig.~\ref{fig:m2_linefit}) appear unrelated to what is seen in J- and K-band. We can identify at least two kinematic components in each object, and it is clear that the Keplerian line profile is absent. Out of the four objects for which we have data in both K- and M-band, the $v \sin i$ is higher in M-band for two of the objects and significantly lower for the other two. 

%% LaTeX deluxetable generator for the AASTeX package.
%% Written by Greg Schwarz (5/1/2001).

%% Table generated: Sun Dec 28 20:30:23 2025

%% Remove the two lines and the last line if you want
%% want to incorporate this table into another LaTex document.

%% The values (usually only l,r and c) in the last part of
%% \begin{deluxetable}{} command tell LaTeX how many columns
%% there are and how to align them.
\renewcommand{\arraystretch}{1.3}
\begin{deluxetable*}{lccccc}[h]

%% Keep a portrait orientation

%% Over-ride the default font size
%% Use Default (12pt)

%% Use \tablewidth{?pt} to over-ride the default table width.
%% If you are unhappy with the default look at the end of the
%% *.log file to see what the default was set at before adjusting
%% this value.

%% This is the title of the table.
\label{tab:vsini}
\tablecaption{$v\sin i$ of FUors and FUor-like objects. References are listed for optical $v \sin i$. The breakup velocity is estimated according to the process in Sec.~\ref{sec:results}.}

%% This command over-rides LaTeX's natural table count
%% and replaces it with this number.  LaTeX will increment 
%% all other tables after this table based on this number

%% The \tablehead gives provides the column headers.  It
%% is currently set up so that the column labels are on the
%% top line and the units surrounded by ()s are in the 
%% bottom line.  You may add more header information by writing
%% another line between these lines. For each column that requries
%% extra information be sure to include a \colhead{text} command
%% and remember to end any extra lines with \\ and include the 
%% correct number of &s.
\tablehead{
\colhead{Object} & \colhead{Optical} & \colhead{J-band} & \colhead{K-band} & \colhead{M-band\tablenotemark{b}} & \colhead{Optical Reference} \\
\colhead{} & \colhead{(km/s)} & \colhead{(km/s)} & \colhead{(km/s)} & \colhead{(km/s)} & \colhead{}
}
%% All data must appear between the \startdata and \enddata commands
\startdata
V2775 Ori       &    &                            &  $5.8^{+0.017}_{-0.016}$ &  &  \\
V1515 Cyg       & 15 &  $7.7^{+0.23}_{-0.23}$     &  $9.3^{+0.034}_{-0.034}$ & $12.7^{+0.12}_{-0.12}$ & \cite{2022ApJ...936...64S} \\
V1735 Cyg       &    &  $11^{+0.41}_{-0.31}$      &  $15^{+0.024}_{-0.025}$ & $21^{+0.10}_{-0.10}$ &  \\
V733 Cep        &    &  $9.1^{+0.39}_{-0.36}$     &  $26^{+0.47}_{-0.35}$ &  &  \\
V2494 Cyg       &    &                            &  $30^{+0.23}_{-0.18}$ & $5.7^{+0.040}_{-0.040}$ &  \\
V960 Mon        & 44 &                            &  $32^{+0.058}_{-0.058}$ &  & \cite{2020ApJ...900...36P} \\
RNO 1b          &    &  $37^{+1.3}_{-1.2}$        &  $33^{+0.12}_{-0.23}$ &  &  \\
V646 Pup        &    &  $55^{+1.7}_{-1.6}$        &  $35^{+0.26}_{-0.26}$ &  &  \\
V900 Mon        &    &                            &  $45^{+0.41}_{-0.35}$ &  &  \\
FU Ori          & 70 &  $50^{+1.5}_{-1.4}$        &  $44^{+0.47}_{-0.35}$ & $16^{+0.088}_{-0.10}$ &
\makecell{\cite{2003ApJ...595..384H} \\ \cite{2008AJ....136..676P}} \\
Parsamian 21    &    &  $35^{+1.9}_{-1.8}$        &  $46^{+0.25}_{-0.25}$ &  &  \\
Haro 5a/6a      &    &                            &  $48^{+0.61}_{-0.61}$ &  &  \\
V883 Ori        &    &  $64^{+1.2}_{-2.1}$        &  $55^{+0.35}_{-0.41}$ & $6.1^{+0.016}_{-0.016}$ &  \\
V582 Aur        &    &                            &  $60^{+1.4}_{-1.2}$  &  &  \\
%PR Ori B\tablenotemark{a} &    & $192^{+2.2}_{-2.1}$ &  $107^{+0.64}_{-0.63}$ &  &  \\
HBC 722         &    &                            &  & $6.6^{+0.048}_{-0.040}$ & \\
\enddata

%% Include any \tablenotetext{key}{text}, \tablerefs{ref list},
%% or \tablecomments{text} between the \enddata and 
%% \end{deluxetable} commands

%% No \tablecomments indicated
\tablenotetext{a}{The $v\sin i$ of PR Ori B are overestimates because we could not obtain a good fit to its spectrum.}
\tablenotetext{b}{M-band values are measured from the half-width at half-depth of the lines themselves. Conversely, the $v \sin i$ in J- and K-band are the half-width at half-maximum of the line profiles.}

%% No \tablerefs indicated

\end{deluxetable*}

\clearpage
\section{Discussion} \label{sec:discussion}
There are a several reasons why a double-peaked line profile may not readily be seen in the data. First, the continuum is significantly contaminated by other absorption features that blend with the CO lines. FU Ori exhibits mores box-like CO lines at first glance, but their shape is reproduced by a double-peaked line profile when a suitable template spectrum is selected. This has also been noticed by \cite{2004ApJ...609..906H}, who showed that using V1515 Cyg's spectrum as a template produced fewer double peaked-CO lines compared to a theoretical model when convolved with a Keplerian profile.

Furthermore, the lack of a clear wavelength dependence with linewidth does not necessarily mean that the Keplerian disk model is incorrect. This effect has been noticed by others in optical spectra, including \cite{2008AJ....136..676P} and \cite{2023ApJ...958..140C}. Changing the maximum temperature of the disk model can eliminate the wavelength dependence at optical wavelengths \citep{2009ApJ...694L..64Z}.

Although there is overlap, we also notice that objects with larger $v \sin i$ are more likely to exhibit double-peaked line profiles. In K-band, the mean $v \sin i$ is 46 $\pm$ \qty{0.3}{\kilo\meter\per\second} for the double-peaked and 36 $\pm$ \qty{0.1}{\kilo\meter\per\second} for the single-peaked line profiles. This implies that inclination is a contributing factor. Additionally, the objects with higher $v\sin i$ that are statistically matched with single-peaked line profiles (Haro 5a and Parsamian 21) have somewhat lower $S/N$ spectra, which renders it more difficult to discern the line structure in these objects.

\begin{figure}[htpb]
    \centering
    \includegraphics[width=\linewidth]{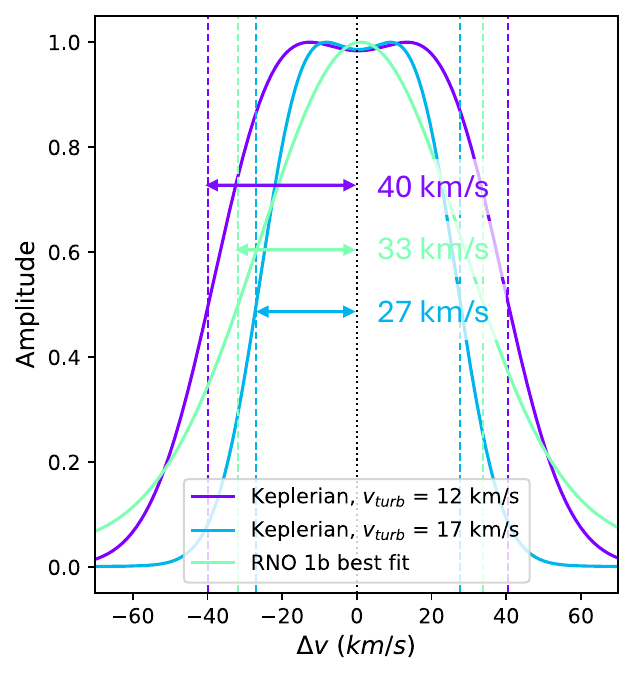}
    \caption{Keplerian line profiles with different amounts of turbulent broadening compared to the line profile of RNO 1b. Below a $v\sin i$ of \qty{40}{\kilo\meter\per\second}, it becomes increasingly difficult to distinguish the two peaks of the Keplerian profile depending on the amount of turbulent broadening.}
    \label{fig:turbulence}
\end{figure}

We also find that a large 12 to \qty{17}{\kilo\meter\per\second} of Gaussian broadening is necessary to smooth out the Keplerian line profiles in in K-band. This is comparable to the amount of turbulent broadening adopted by \cite{2023ApJ...958..140C} to match the optical spectra of V960 Mon. Figure~\ref{fig:turbulence} shows that if the $v\sin i$ is less than 30 to \qty{40}{\kilo\meter\per\second}, the peaks may become blended from turbulence in the disk. V646 Pup has a larger $v\sin i$ than this and is not fitted by a double-peaked line profile in J- or K-band, although it is a bit atypical in that is CO bandhead is weaker than in other FUors (also see Appendix~\ref{sec:prori}).

\subsection{Excess absorption--a disk wind?}

\begin{figure}[htpb]
    \centering
    \includegraphics[width=\linewidth]{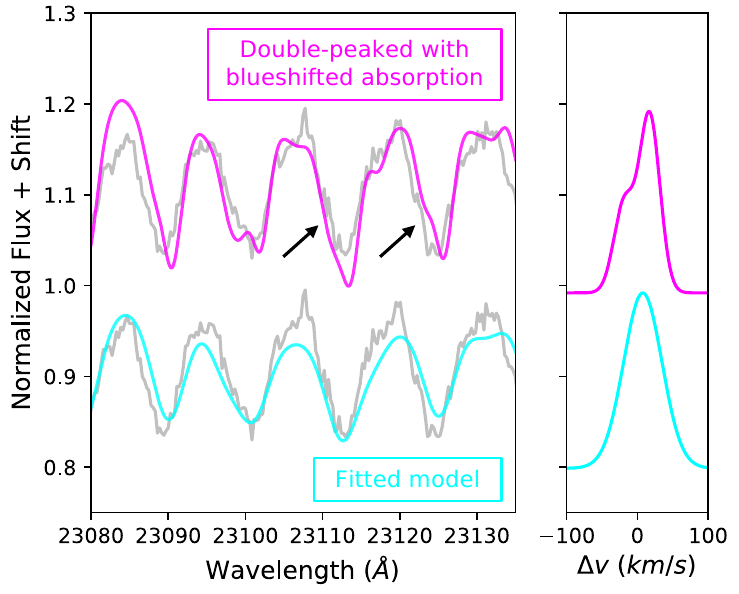}
    \caption{CO lines of RNO 1b (gray) compared to the fitted single-peaked line profile (cyan) and a test model with a stronger blueshifted component of absorption (pink). Although the overall fit of the test model is worse, it is better at reproducing the slightly bowed, triangular shape of the lines (next to the arrows) and steeper slope between each line.}
    \label{fig:rno1b-wind}
\end{figure}

\begin{figure*}[htpb]
    \centering
    \includegraphics[width=0.8\linewidth]{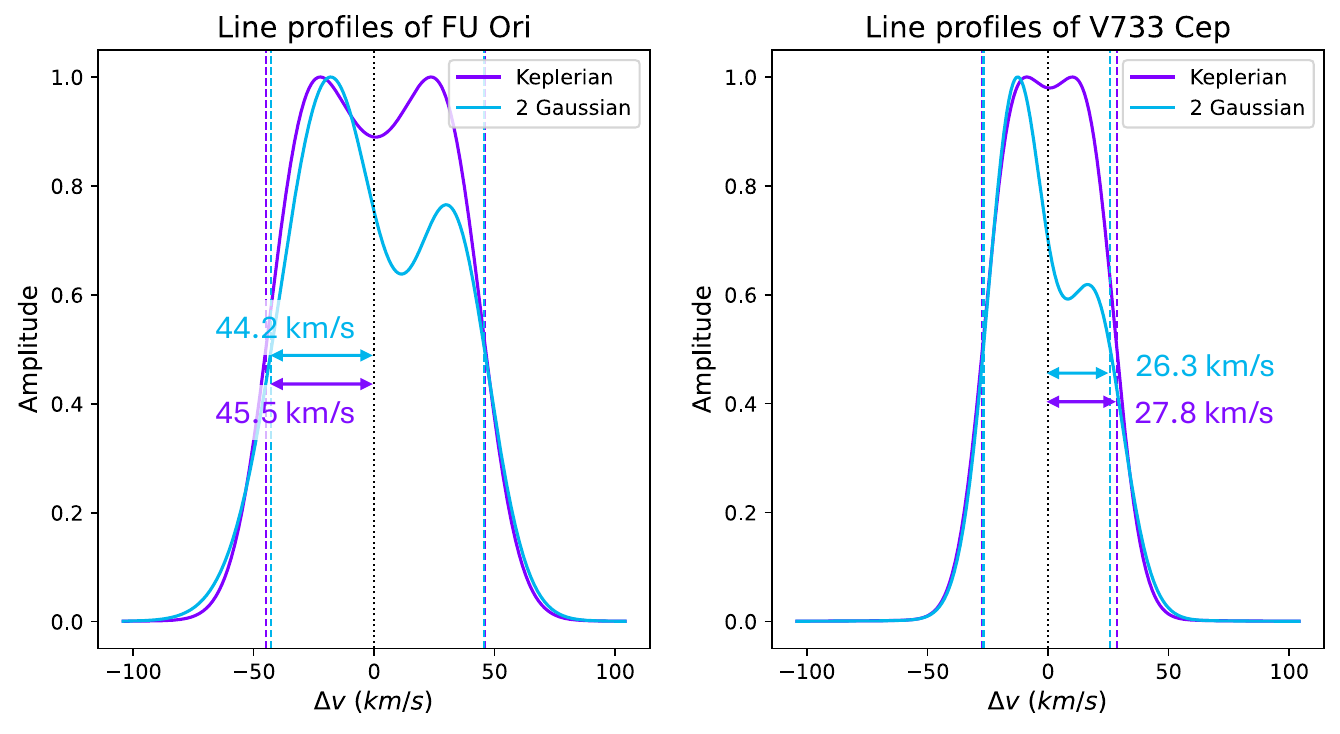}
    \caption{A comparison of the HWHM measured for symmetrical Keplerian line profiles versus a linear combination of two gaussians for two objects with asymmetrical line profiles. The HWHM measures the rotational velocities fairly consistently despite the asymmetry, contributing approximately \qty{1.5}{\kilo\meter\per\second} of uncertainty. The line profiles are normalized for visualization (a constant scaling factor does not affect the HWHM).}
    \label{fig:hwhm}
\end{figure*}

The double-peaked line profiles may also be suppressed by excess absorption. For example, Fig.~\ref{fig:rno1b-wind} shows that several features in the CO lines of RNO 1b are better described by a double-peaked line profile with an additional blueshifted component of absorption, which has previously been attributed to a disk wind. This would not severely degrade our $v \sin i$ measurements given that its width is confined to that of the disk model \citep{2024ApJ...971...44C}. Figure~\ref{fig:hwhm} shows that the HWHM is affected by less than \qty{2}{\kilo\meter\per\second} even when there is a large excess component of absorption.

That being said, the differences between single-peaked and asymmetrical double-peaked line profiles are subtle; it is difficult to tell what instead may be caused a deficiency in the template spectrum. It is apparent that the CO lines of FUors exhibit a more complex line profile than predicted by a basic disk model, as opposed to the Al lines in J-band which are usually well fitted by a symmetrical Keplerian line profile if they appear to be double-peaked.

\begin{figure}[htpb]
    \centering
    \includegraphics[width=\linewidth]{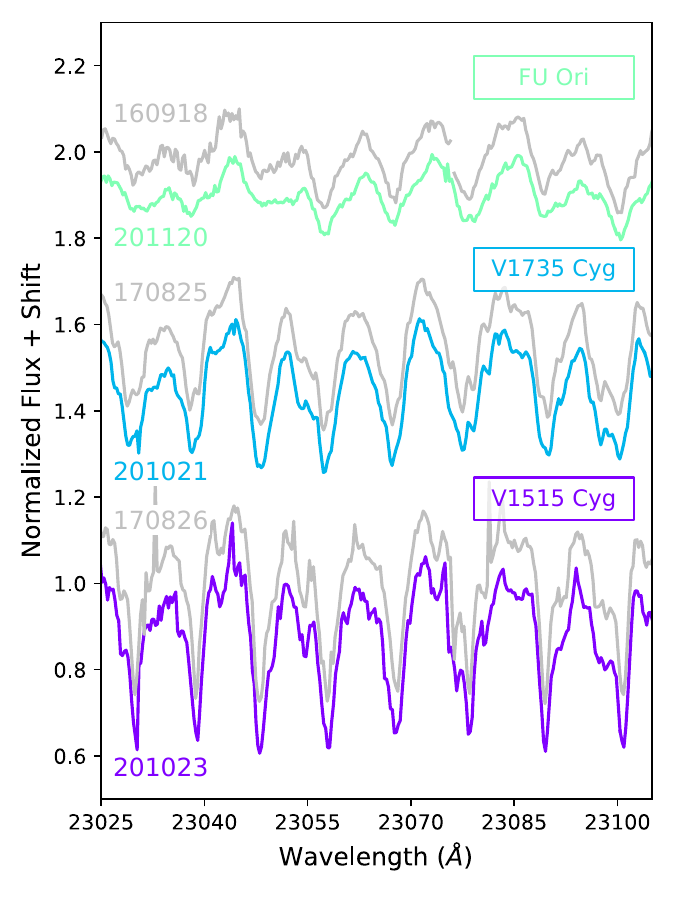}
    \caption{Comparison of CO lines for objects observed in K-band on multiple dates. The spectra have not changed much in three to four years.}
    \label{fig:variable-k2}
\end{figure}

We can also examine the few objects for which we have collected data over multiple years. Figure~\ref{fig:variable-k2} shows that the CO line structure is quite stable, although our baseline is only four years. If a CO wind is responsible for the excess absorption, it exhibits significantly less variability than the winds that are known to impact H$\alpha$, Ca II, and other lines that are typically used to trace winds in young stars. These lines are known to vary on timescales of days in FUors and show more extended blueshifted absorption than we see in the CO lines \citep{herbig2003high, 2023ApJ...958..140C}.

\subsection{Comparisons to the expected Keplerian velocity}

It is possible to estimate the size of the emitting region by assuming a blackbody temperature of \qty{3000}{\kelvin} and using the K-band magnitudes $(\lambda \approx \qty{2.2}{\micro\meter})$ and distances. The radius is roughly 
\begin{equation}
    R \approx d \sqrt{\frac{f_{\lambda,recv}}{f_{\lambda,emit}}}
\end{equation}
where $d$ is the distance, $f_{\lambda,recv}$ is the received flux density calculated from the magnitude, and $f_{\lambda,emit}$ is the flux density emitted by a \qty{3000}{\kelvin} blackbody.  The Keplerian velocity is
\begin{equation}
    v_K = \sqrt{\frac{GM}{R}}
\end{equation}
for which we can use $1 M_{\odot}$ as the mass of the central star if it is unknown. The Keplerian velocities are provided in Table~\ref{tab:breakup}.

One would expect the deprojected rotational velocities ($v_{rad}$) to be close to the Keplerian velocity. We can compare a few objects for which measurements of the stellar mass and inclination from ALMA data exist. Using a \qty{38.1}{\degree} inclination and a stellar mass of $1.3 M_{\odot}$ for V883 Ori \citep{2016Natur.535..258C}, its $v_{rad}$ is \qty{89}{\kilo\meter\per\second} which is close to the approximated Keplerian velocity. The inclination of FU Ori is suggested to be around \qty{37.7}{\degree} and its stellar mass $0.6 M_{\odot}$ \citep{2020ApJ...889...59P}, meaning that its $v_{rad} = \qty{72}{\kilo\meter\per\second}$. This is substantially larger than the Keplerian velocity, but near-infrared interferometric measurements of the inclination are closer to \qty{55}{\degree} due to misalignment of the inner disk \citep{2005A&A...437..627M, lykou2022disk}. This yields $v_{rad} = \qty{54}{\kilo\meter\per\second}$, which is much closer to the Keplerian velocity.

\begin{figure}[h]
    \centering
    \includegraphics[width=\linewidth]{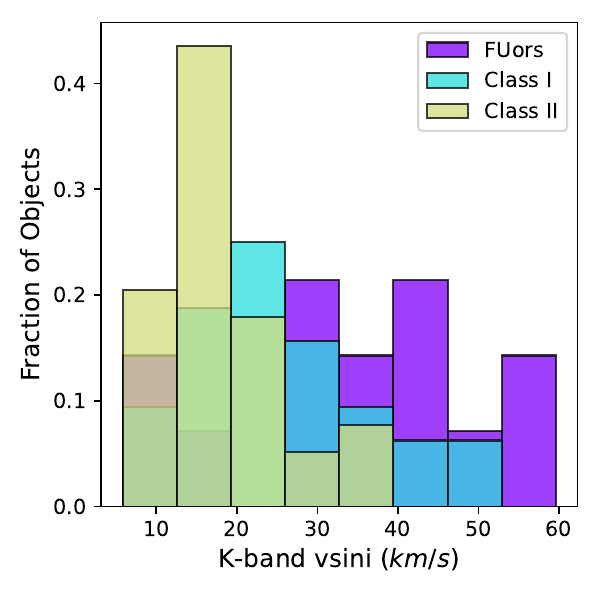}
    \caption{Distribution of $v\sin i$ of the observed FUors $(N=15)$ compared to Class I $(N=32)$ and Class II $(N=39)$ stars measured by \cite{2024ApJ...972..149F}. The $v\sin i$ for our sample is comparable to those measured for Class I objects.}
    \label{fig:vsini_hist}
\end{figure}

Note that our approximation assumes that the emitting region is a sphere. The distribution of K-band $v\sin i$ for the FUors is comparable to that of Class I objects (Fig.~\ref{fig:vsini_hist}), which is consistent with the fact that the calculated $R$ are similar to the co-rotation radii of T Tauri stars \citep{pittman2025odysseus}. The Keplerian velocities are therefore approximated fairly well.

%% LaTeX deluxetable for ApJ (AASTeX)
\renewcommand{\arraystretch}{1.3}
\begin{deluxetable}{lcccc}[h]

\label{tab:breakup}
\tablecaption{K-band $v\sin i$, distance, characteristic radius $R$, and Keplerian velocity for FUors and FUor-like objects.}

\tablehead{
\colhead{Object} &
\colhead{K-band $v\sin i$} &
\colhead{Distance\tablenotemark{a}} &
\colhead{$R$} &
\colhead{Keplerian Velocity\tablenotemark{b}} \\
\colhead{} &
\colhead{(km s$^{-1}$)} &
\colhead{(pc)} &
\colhead{(AU)} &
\colhead{(km s$^{-1}$)}
}

\startdata
V2775 Ori        & $5.8^{+0.017}_{-0.016}$   & 446 & 0.020  & 212    \\
V1515 Cyg        & $9.3^{+0.034}_{-0.034}$   & 960 & 0.14   & 79    \\
V1735 Cyg        & $15^{+0.024}_{-0.025}$    & 752 & 0.13   & 82    \\
V733 Cep         & $26^{+0.47}_{-0.35}$      & 825 & 0.084  & 103    \\
V2494 Cyg        & $30^{+0.23}_{-0.18}$      & 594 & 0.058  & 124    \\
V960 Mon         & $32^{+0.058}_{-0.058}$    & 1120 & 0.11  & 90     \\
RNO 1b           & $33^{+0.12}_{-0.23}$      & 930 (1) & 0.11 & 88    \\
V646 Pup         & $35^{+0.26}_{-0.26}$      & 1800 (2) & 0.22   & 63    \\
V900 Mon         & $45^{+0.41}_{-0.35}$      & 1120 & 0.11   & 90   \\
FU Ori           & $44^{+0.47}_{-0.35}$      & 404 & 0.16   & 57 (6)   \\
Parsamian 21     & $46^{+0.25}_{-0.25}$      & 500 (3) & 0.024  & 191     \\
Haro 5a/6a       & $48^{+0.61}_{-0.61}$      & 388 (4) &  0.018 & 222      \\
V883 Ori         & $55^{+0.35}_{-0.41}$      & 392 & 0.28   & 85 (7)     \\
V582 Aur         & $60^{+1.4}_{-1.2}$        & 1300 (5) & 0.17   & 73     \\
%PR Ori B         & $107^{+0.64}_{-0.63}$     & 331* & 0.034  & 230    \\
\enddata
\tablenotetext{a}{When available, the distances are taken from \cite{2019ApJ...883..117K}. Otherwise, we use distances from the literature found by \cite{connelley2018fuor}. Specifically: (1) \cite{reid2014trigonometric}(2) \cite{reipurth2002evolution} (3) \cite{dame1984wide} (4) \cite{kounkel2017gould} (5) \cite{kun2017star}}
\tablenotetext{b}{We assume $1 M_{\odot}$ unless a measurement for the mass exists: (6) \cite{2020ApJ...889...59P} (7) \cite{2016Natur.535..258C}}

\end{deluxetable}

\section{Conclusions} \label{sec:conclusions}
We have shown that the near-infrared line profiles of FUors and FUor-like objects may not always appear to be Keplerian, but this does not necessarily mean that the light is not coming from a disk. The line profiles can be affected by a variety of factors and would require additional components on top of a basic accretion disk model to fully explain. In summary:
\begin{enumerate}
    \item The 15 objects observed at K-band are well-described by either single- or double-peaked line profiles. The line profiles seem to be stable over baselines of three to four years.
    \item The double-peaked line profiles can be difficult to see in the CO lines due to blending with other features in the spectrum. There may also be blueshifted or redshifted components of absorption that result in one of the peaks becoming much deeper, which further causes the lines to appear single-peaked.
    \item For the eight objects that were also observed in J-band and for which we obtained a reasonable fit, we do not see a trend of decreasing linewidth at redder wavelengths. This is expected from a basic Keplerian disk model, but the lack of a clear wavelength dependence has been predicted by more sophisticated models and observed by others at optical wavelengths.
    \item The Keplerian line profile is not observed in the CO lines in M-band. The structure of these lines appears unrelated to what is observed in J- and K-band, so they likely do not originate from the disk.
\end{enumerate}

\section*{Acknowledgments}
We would like to thank Lynne Hillenbrand and the reviewer for their helpful feedback regarding the analysis in this paper. The authors of this work are funded by the Infrared Telescope Facility, which is operated by the University of Hawaii under contract 80HQTR24DA010 with the National Aeronautics and Space Administration. This research has made use of the VizieR catalogue access tool, CDS, Strasbourg, France \citep{10.26093/cds/vizier}. The original description ßof the VizieR service was published in \citet{vizier2000}.

\appendix

\section{PR Ori B} \label{sec:prori}
\begin{figure}
    \centering
    \includegraphics[width=0.85\linewidth]{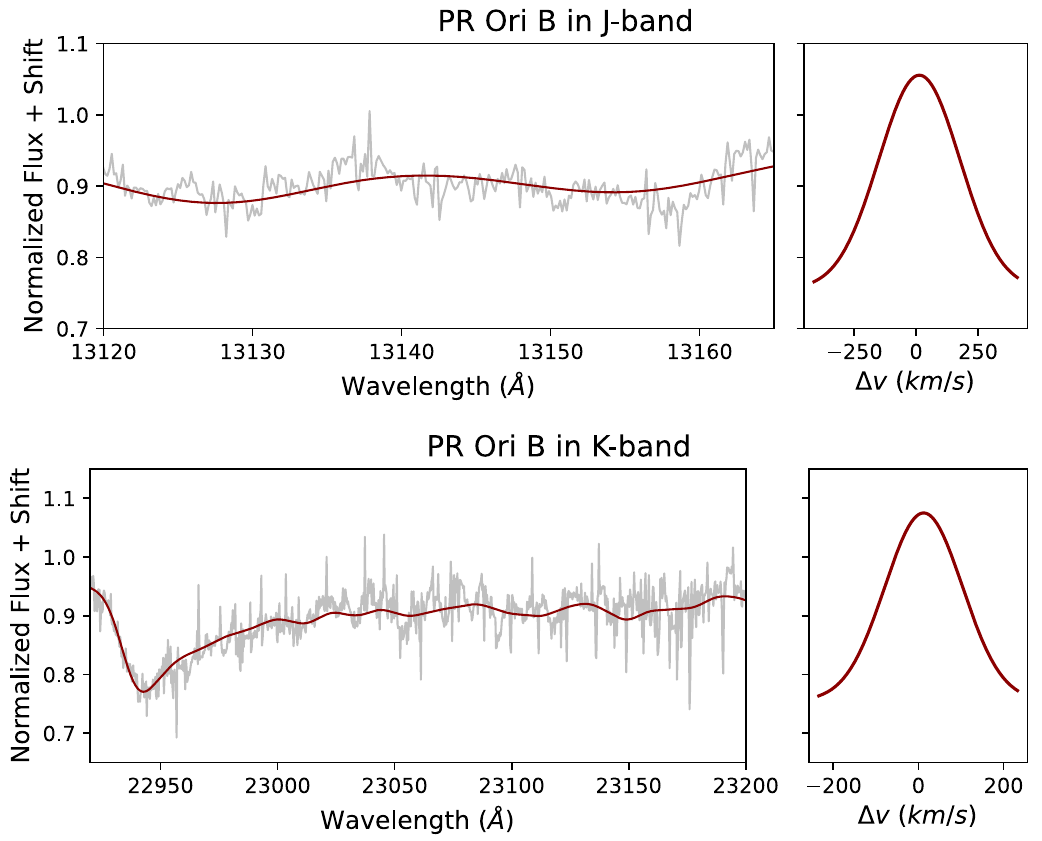}
    \caption{Best-fit line profiles for PR Ori B. Our models are a poor fit to the spectra, particularly in K-band where there are narrower CO lines on top of the very highly broadened spectrum.}
    \label{fig:prori}
\end{figure}

PR Ori B is separated from the rest of the objects in our sample as we find that its spectrum is not well-described by our models. Based on the line profiles shown in Fig.~\ref{fig:prori}, we find a $v \sin i$ of $192^{+2.2}_{-2.1}$ \qty{}{\kilo\meter\per\second} in J-band and $107^{+0.64}_{-0.63}$ \qty{}{\kilo\meter\per\second} in K-band. However, it is clear that data are fitted poorly and that the $v \sin i$ is overestimated in J-band. We were unable to obtain a reasonable fit to the spectrum by adding more Gaussians to the line profile, so the templates may not be suitable for this object.

The bandhead of PR Ori B indicates a significantly larger $v\sin i$ than the rest of the CO band. We notice a similar effect in a few other objects in our sample, albeit to a lesser degree, shown in Fig.~\ref{fig:bandhead}. This was also seen by \cite{2020ApJ...900...36P} for V960 Mon. Rather than different amounts of rotational broadening, it is likely that the bandhead itself is intrinsically weaker in FUors; a progressively weakening bandhead is also observed in V1057 Cyg as its eruption decays \citep{2021ApJ...917...80S}. However, PR Ori B is further distinguished from the rest of the FUors by the fact that its bandhead lacks a shoulder, which is expected for a rotating disk \citep{carr1993inner}. We clearly see this feature in the FUors with large $v \sin i$ that are fitted by double-peaked line profiles (and arguably in Parsamian 21 as well). Thus, the classification of PR Ori B as a FUor is somewhat questionable when combined with the very large $v \sin i$, which is more indicative of PR Ori B being a Herbig Ae/Be star.

\begin{figure}
\renewcommand{\theHfigure}{\arabic{figure}a}
    \centering
    \includegraphics[width=\linewidth]{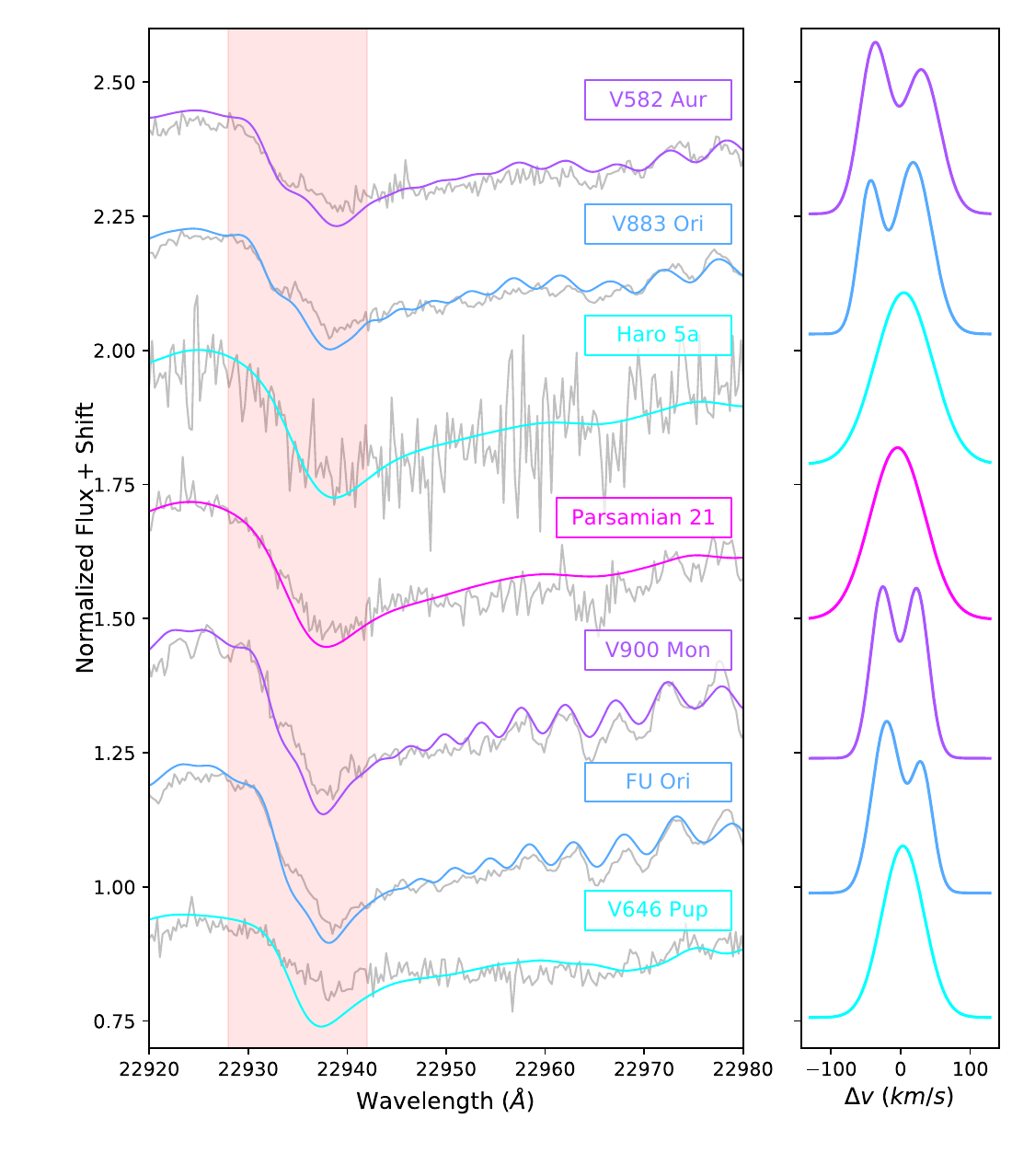}
    \caption{Zoomed in region of the CO bandhead for the models shown in Fig.~\ref{fig:spectra_k2}. The template spectra appear to consistently overestimate the strength of the bandhead (shaded in red).}
    \label{fig:bandhead}
\end{figure}

\begin{figure}
    \centering
    \addtocounter{figure}{-1}
    \renewcommand{\theHfigure}{\arabic{figure}b}
    \includegraphics[width=\linewidth]{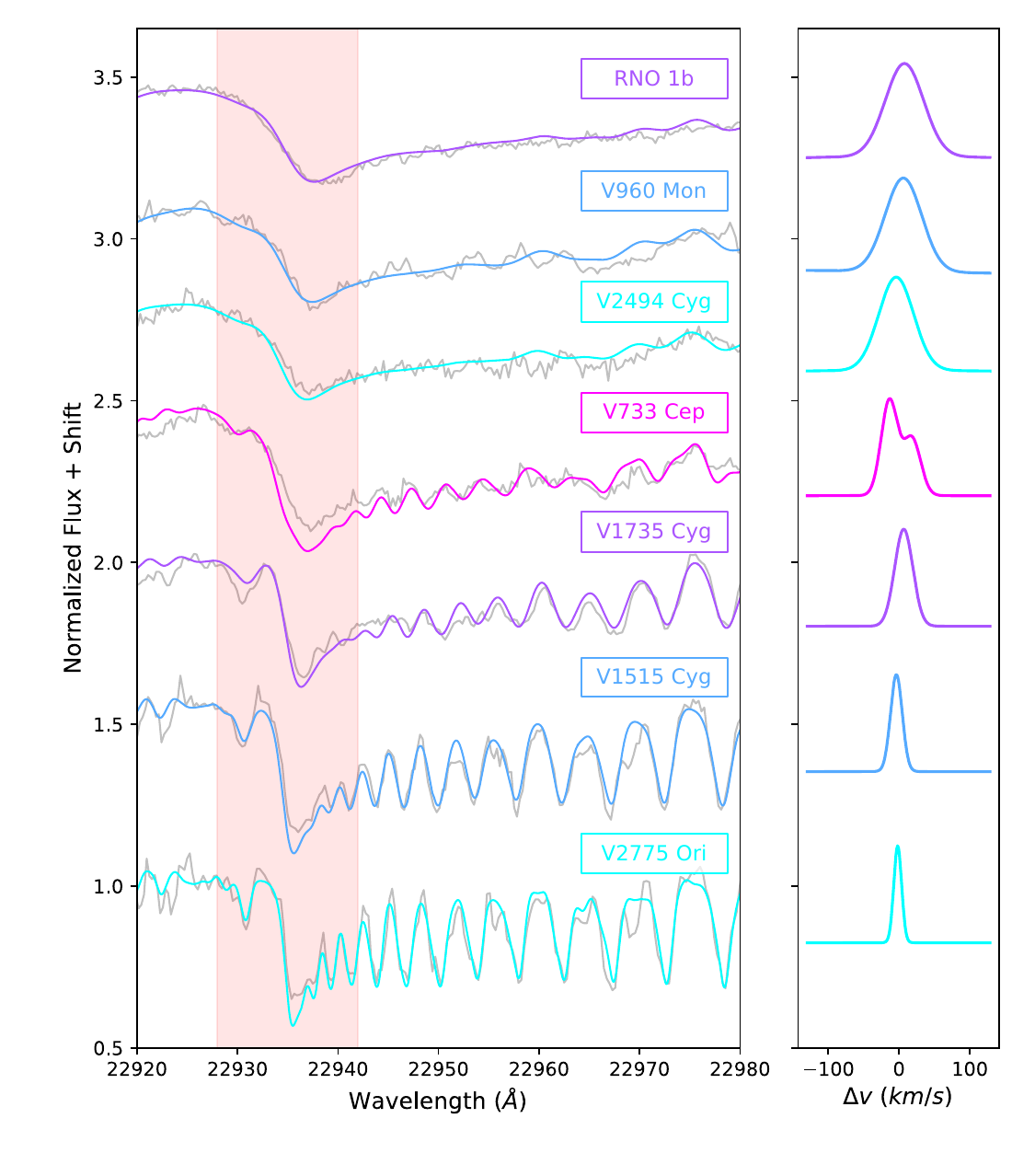}
    \caption{Continued.}
\end{figure}

\bibliography{paper}{}
\bibliographystyle{aasjournal}

\end{document}